\documentclass[a4paper,11pt,english]{amsart}
\usepackage[left=25mm,top=30mm,bottom=25mm,right=25mm]{geometry}
\usepackage{amssymb,dsfont,bm}
\usepackage[utf8]{inputenx} 
\usepackage{babel}
\usepackage{cmap}
\usepackage{hyperxmp}
\usepackage{graphicx}
\usepackage[pdfdisplaydoctitle=true,
      colorlinks=true,
      urlcolor=blue,
      citecolor=blue,
      linkcolor=blue,
      pdfstartview=FitH,
      pdfpagemode=UseNone,
      bookmarksnumbered=true]{hyperref} 
\usepackage{tikz}
\usetikzlibrary{arrows}

\graphicspath{{figs/}}
\newtheorem{thm}{Theorem}[section]

\newtheorem{lem}[thm]{Lemma}

\theoremstyle{definition}

\theoremstyle{remark}
\newtheorem{proc}[thm]{Procedure}
\newtheorem{rem}[thm]{Remark}
\newtheorem{ex}{Example}[section]

\numberwithin{equation}{section}

\newcommand{\RR}{\mathbb{R}}                
\newcommand{\ZZ}{\mathbb{Z}}                

\newcommand{\Ela}{\mathbb{E}\mathrm{la}}    

\newcommand{\OO}{\mathrm{O}}                
\newcommand{\SO}{\mathrm{SO}}               
\newcommand{\octa}{\mathbb{O}}              
\newcommand{\DD}{\mathbb{D}}                
\newcommand{\triv}{\mathds{1}}		        

\newcommand{\ee}{\pmb{e}}                   
\newcommand{\nn}{\pmb{n}}                   
\newcommand{\uu}{\pmb{u}}                   
\newcommand{\vv}{\pmb{v}}                   
\newcommand{\ww}{\pmb{w}}                   

\newcommand{\Idd}{\mathbf{1}}               
\newcommand{\ba}{\mathbf{a}}
\newcommand{\bb}{\mathbf{b}}
\newcommand{\bc}{\mathbf{c}}
\newcommand{\bd}{\mathbf{d}}

\newcommand{\bv}{\mathbf{v}}
\newcommand{\bt}{\mathbf{t}}

\newcommand{\bomega}{\bm{\omega}}
\newcommand{\bepsilon}{\bm{\varepsilon}}

\newcommand{\bJ}{\mathbf{J}}                
\newcommand{\bE}{\mathbf{E}}                
\newcommand{\bH}{\mathbf{H}}                
\newcommand{\bS}{\mathbf{S}}                
\newcommand{\bC}{\mathbf{C}}                
\newcommand{\bA}{\mathbf{A}}                
\newcommand{\bB}{\mathbf{B}}                

\DeclareMathOperator{\tr}{tr}
\newcommand{\rot}{\mathbf{r}}

\newcommand{\threedots}{\mathbin{\raisebox{-0.25ex}{\vdots}}}
\newcommand{\otimesbar}{\mathbin{\underline{\overline{\otimes}}}}

\newcommand{\norm}[1]{\lVert#1\rVert}       
\newcommand{\set}[1]{\left\{#1\right\}}     

\hypersetup{
  pdfauthor={S. Abramian, B. Desmorat, B. Kolev, M. Olive and R. Desmorat}, 
  pdftitle={Recovering the normal form and symmetry class of an Elasticity tensor}, 
  pdfsubject={MSC 2010: 74E10 (15A72 74B05)}, 
  pdfkeywords={Elasticity tensor; Natural basis; Symmetry classes; Covariants}, 
  pdflang=en 
 }
\begin{document}

\title[Recovering normal forms]{Recovering the normal form and symmetry class of an Elasticity tensor}

\author{S. Abramian}
\address[Sophie Abramian]{Université Paris-Saclay, ENS Paris-Saclay, CNRS, LMT - Laboratoire de Mécanique et Technologie, 91190, Gif-sur-Yvette, France}
\email{sophie.abramian@ens-paris-saclay.fr}

\author{B. Desmorat}
\address[Boris Desmorat]{Sorbonne Université, UMPC Univ Paris 06, CNRS, UMR 7190, Institut d'Alembert, F-75252 Paris Cedex 05, France \& Univ Paris Sud 11, F-91405 Orsay, France}
\email{boris.desmorat@sorbonne-universite.fr}

\author{R. Desmorat}
\address[Rodrigue Desmorat]{Université Paris-Saclay, ENS Paris-Saclay, CNRS, LMT - Laboratoire de Mécanique et Technologie, 91190, Gif-sur-Yvette, France}
\email{rodrigue.desmorat@ens-paris-saclay.fr}

\author{B. Kolev}
\address[Boris Kolev]{Université Paris-Saclay, ENS Paris-Saclay, CNRS, LMT - Laboratoire de Mécanique et Technologie, 91190, Gif-sur-Yvette, France}
\email{boris.kolev@math.cnrs.fr}

\author{M. Olive}
\address[Marc Olive]{Université Paris-Saclay, ENS Paris-Saclay, CNRS, LMT - Laboratoire de Mécanique et Technologie, 91190, Gif-sur-Yvette, France}
\email{marc.olive@math.cnrs.fr}

\thanks{R. Desmorat, B. Kolev and M.Olive were partially supported by CNRS Projet 80$|$Prime GAMM (Géométrie algébrique complexe/réelle et mécanique des matériaux)}%

\subjclass[2010]{74E10 (15A72 74B05)}%
\keywords{Elasticity tensor; Anisotropy; Natural basis; Symmetry classes; Covariants}%

\date{June 22, 2020}%
\begin{abstract}
  We propose an effective geometrical approach to recover the normal form of a given Elasticity tensor. We produce a rotation which brings an Elasticity tensor onto its normal form, given its components in any orthonormal frame, and this for any tensor of any symmetry class. Our methodology relies on the use of specific covariants and on the geometric characterization of each symmetry class using these covariants. An algorithm to detect the symmetry class of an Elasticity tensor is finally formulated.
\end{abstract}

\maketitle
\section{Introduction}
\label{sec:intro}

The linear elastic properties of a given material are encoded into an Elasticity tensor $\bE$, a fourth-order tensor which relates linearly the stress tensor  to the strain tensor. As it was clearly emphasized by Boehler and coworkers~\cite{Boe1987,BKO1994}, any rotated Elasticity tensor encodes the same material properties (in a different orientation). One shall say that the rotated tensor and initial one are in the same orbit.
It should be emphasized here that this has not to be confused with a change of (orthonormal) basis once a basis has been fixed and the tensors expressed by their components in this basis. Here, the action of the rotation group is defined intrinsically and independently of any basis (no components are required to define this action).

The elastic materials are classified by their eight symmetry classes~\cite{FV1996} (isotropic, transversely-isotropic, cubic, trigonal, tetragonal, orthotropic, monoclinic, triclinic). Any non triclinic Elasticity tensor has a given symmetry class and a \emph{normal form}. An orthonormal frame in which the matrix representation of this tensor belongs to such a normal form is called a \emph{proper or natural basis} for $\bE$~\cite{Fed1968}. For instance, consider a cubic Elasticity tensor which is given in an arbitrary frame by its Voigt's (matrix) representation $[\bE]$  (not to be confused with the tensor $\bE$ itself) as
\begin{equation}\label{eq:Voigt}
  [\bE] =
  \begin{pmatrix}
    E_{1111} & E_{1122} & E_{1133} & E_{1123} & E_{1113} & E_{1112} \\
    E_{2211} & E_{2222} & E_{2233} & E_{2223} & E_{2213} & E_{2212} \\
    E_{3311} & E_{3322} & E_{3333} & E_{3323} & E_{3313} & E_{3312} \\
    E_{2311} & E_{2322} & E_{2333} & E_{2323} & E_{2313} & E_{2312} \\
    E_{1311} & E_{1322} & E_{1333} & E_{1323} & E_{1313} & E_{1312} \\
    E_{1211} & E_{1222} & E_{1233} & E_{1223} & E_{1213} & E_{1212}
  \end{pmatrix}.
\end{equation}
Then, there exists a rotation $g$ such that the rotated Elasticity tensor, denoted by $g\star \bE$, and where
\begin{equation}\label{eq:gstarE}
  (g\star \bE)_{ijkl} = g_{ip}g_{jq} g_{kr}g_{ls}E_{pqrs}
\end{equation}
has the following Voigt's representation
\begin{equation}\label{eq:intro-cube}
  [g\star \bE] =
  \begin{pmatrix}
    (g\star \bE)_{1111} & (g\star \bE)_{1112} & (g\star \bE)_{1112} & 0                   & 0                   & 0                  \\
    (g\star \bE)_{1112} & (g\star \bE)_{1111} & (g\star \bE)_{1112} & 0                   & 0                   & 0                  \\
    (g\star \bE)_{1112} & (g\star \bE)_{1112} & (g\star \bE)_{1111} & 0                   & 0                   & 0                  \\
    0                   & 0                   & 0                   & (g\star \bE)_{1212} & 0                   & 0                  \\
    0                   & 0                   & 0                   & 0                   & (g\star \bE)_{1212} & 0                  \\
    0                   & 0                   & 0                   & 0                   & 0                   & (g\star\bE)_{1212}
  \end{pmatrix}.
\end{equation}
The problem is that it is not always easy to compute explicitly such a rotation. For instance, given a cubic Elasticity tensor in its normal form~\eqref{eq:intro-cube} and applying a rotation of angle $\frac{\pi}{6}$ around axis $<111>$, it is not an easy matter, if not aware of this transformation, to find a way back. Moreover, measured tensors are in practice triclinic, due to numerical errors and experimental discrepancy~\cite{Art1993,Fra1995,DG2003,GS2010}. Hence, the problem may also be numerically difficult.

Partial answers concerning the explicit determination of a proper basis have already been investigated in~\cite{CM1987,Cow1989,Jar1994,Bae1998,CVC2001} for the monoclinic and the orthotropic symmetry classes. To do so, the authors construct a basis of eigenvectors for the second-order symmetric tensors that inherit (part) of the symmetry of $\bE$, the dilatation and Voigt's tensors~\cite{CM1987,Cow1989}, defined as
\begin{equation}\label{eq:dv}
  \begin{aligned}
    \bd: = \tr_{12} \bE\qquad (d_{ij} = E_{kkij}), \\
    \bv: = \tr_{13} \bE\qquad (v_{ij} = E_{kikj}).
  \end{aligned}
\end{equation}
The cornerstone of this approach is that $\bd = \bd(\bE)$ and $\bv = \bv(\bE)$ are covariants of $\bE$, meaning that one has the \emph{covariance (equivariance) property}
\begin{equation*}
  \bd(g\star \bE) = g\star \bd(\bE),\qquad \bv(g\star \bE) = g\star \bv(\bE),
\end{equation*}
where $(g\star \ba)_{ij} = g_{ik}g_{jl} a_{kl}$, for a second-order tensor $\ba$. In some non-degenerate cases, this leads to the answer. The weakness of this approach is that $\bd$ and $\bv$ have at least the symmetry of $\bE$ but they may have more symmetry. For instance, in the cubic case, the pair $(\bd, \bv)$ is isotropic. Such loss of information has to be handled, as they can be experimentally encountered, for example from the ultrasonic measurements made on a Ni base single crystal superalloy, close to be cubic \cite{FBG1998} (studied in section \ref{subsec:Ni-superalloy}).

A natural way is to extend the idea of using covariants of $\bE$, which naturally inherit the symmetry of $\bE$, but different from $\bd$ and $\bv$. Note, however, that second-order covariants cannot always encode all the geometric information carried by a fourth-order tensor~\cite{Boe1979,Liu1982} (for example when $\bE$ is cubic). Taking into account this observation, it has been tried by some authors to use the \emph{harmonic factorization}, according to Sylvester's theorem \cite{Syl1909} and Maxwell's multipoles~\cite{ZTL2013}. However, this involves roots' computations of polynomials of degree $4$ and $8$~\cite{Bac1970,Bae1998,BBS2004}, in order to build a set of $8$ unit vectors (Maxwell's multipoles), without any clue of how to organize such data. Besides, Maxwell's multipoles are not, strictly speaking, first-order covariants of $\bE$ and are moreover very sensitive to conditioning.

The main purpose of the present work is to obtain an explicit normal form of an Elasticity tensor $\bE$ which belongs to a given symmetry class and a rotation which brings $\bE$ to it. Note, by the way, that the problem of determining the symmetry class of a given Elasticity tensor $\bE$, using polynomial covariant equations, has already been solved explicitly in~\cite{OKDD2018b}. Of course, our goal can be achieved numerically, as we can compute $\overline{\bE} = g\star \bE$ for all $g\in \SO(3)$ and try to find rotation $g$ such that Voigt's representation $[\overline{\bE}]$ has the good shape \cite{Fra1995,FBG1998}. A more geometrical approach, initiated in \cite{CM1987,Cow1989,Jar1994,Bae1998}, relying on \emph{covariants}, is possible and will be described in this work. We shall first formulate new effective and fast procedures to calculate a natural basis for a given Elasticity tensor, once we know its symmetry class.

An important tool, introduced in~\cite{DADKO2019,OKDD2018b} and which will be used many times in this paper, is the \emph{generalized cross product} between two \emph{totally symmetric tensors} of any order $\bA = \bA^s$ and $\bB = \bB^s$. It is defined as follows
\begin{equation}\label{eq:cross-product}
  \bA \times \bB = \left(\bB \cdot\pmb \varepsilon \cdot \bA \right)^{s} = -\bB \times \bA,
\end{equation}
where $(\cdot)^s$ means the total symmetrization (over all subscripts) and where $\pmb \varepsilon$ is Levi-Civita third order tensor ($\varepsilon_{ijk} = \det(\ee_i, \ee_j, \ee_k)$ in any direct orthonormal basis $(\ee_{i})$). The expressions of the components of \eqref{eq:cross-product} are given in  \autoref{sec:recovering-normal-forms} and \autoref{sec:detailed-components} for $\bA$ and $\bB$ of order two.

The outline of the paper is as follows. We first recall some mathematical materials on the normal form of an Elasticity tensor in~\autoref{sec:normal-forms} and the harmonic decomposition and the notion of covariants in~\autoref{sec:covariants}. Then, in~\autoref{sec:recovering-normal-forms}, we formulate and prove theorems that are the cornerstones to build our algorithms. In~\autoref{sec:experiment}, we provide and analyze experimental data, issued from the literature and which are used to illustrate our methodology. In~\autoref{sec:effective-computations}, we use our procedures to produce, for any given Elasticity tensor $\bE$, a natural basis for it (and a rotation which brings it back to its normal form). Finally, an algorithm to detect the symmetry class and recover a normal form is detailed in~\autoref{sec:algo}.

\section{Normal form of an Elasticity tensor}
\label{sec:normal-forms}

An Elasticity tensor $\bE$ represents a material in a specific orientation, but the \emph{same material} is represented in another orientation by a rotated tensor $g \star \bE$. In mathematical terms, this means that the rotation group $\SO(3)$ acts linearly on the space $\Ela$ of Elasticity tensors, which we write as
\begin{equation*}
  \bE \mapsto \overline{\bE} = g \star \bE,
\end{equation*}
where
\begin{equation*}
  \overline{E}_{ijkl} = g_{ip}g_{jq} g_{kr}g_{ls}E_{pqrs},
\end{equation*}
in any orthonormal basis $(\ee_{1},\ee_{2},\ee_{3})$. The subset
\begin{equation*}
  \set{g \star \bE; \; g \in \SO(3)}
\end{equation*}
is called the \emph{orbit} of $\bE$. A linear elastic material is thus represented by an orbit of an Elasticity tensor rather than by a specific Elasticity tensor in its orbit.

The \emph{symmetry group} of a tensor $\bE \in \Ela$ is the subgroup of $\SO(3)$ defined as
\begin{equation*}
  G_{\bE}: = \set{g \in \SO(3);\; g \star \bE = \bE}.
\end{equation*}
Note that the symmetry group of $g\star \bE$ is
\begin{equation}
  G_{\overline{\bE}} = g G_{\bE} g^{-1}.
\end{equation}
Therefore, the classification of \emph{symmetries of materials} relies on the \emph{conjugacy classes}
\begin{equation*}
  [G_{\bE}]: = \set{gG_{\bE}g^{-1},\quad g\in \SO(3)},
\end{equation*}
rather than on the symmetry groups of their respective tensors in a specific orientation. These are known as \emph{symmetry classes}.

It was shown in~\cite{FV1996} that there are exactly eight Elasticity symmetry classes: triclinic $[\triv]$, monoclinic $[\ZZ_{2}]$, orthotropic $[\DD_{2}]$, tetragonal $[\DD_{4}]$, trigonal $[\DD_{3}]$, transversely-isotropic $[\OO(2)]$, cubic $[\octa]$ and isotropic $[\SO(3)]$. Each symmetry class is a conjugacy class of subgroups and for each of it, it is useful to \emph{fix} a particular representative subgroup in it. Such representatives are provided, for each Elasticity class, in the list below.

\begin{itemize}
  \item The trivial group is designed by $\triv := \set{I}$, no specific representative is required for it;
  \item $\ZZ_{2}$ is generated by the second-order rotation $\rot(\ee_{3},\pi)$ and is of order $2$;
  \item $\DD_{2}$ is generated by the second-order rotations $\rot(\ee_{3},\pi)$ and $\rot(\ee_{1},\pi)$ and is of order $4$;
  \item $\DD_{3}$ is generated by the third order rotation $\rot(\ee_{3},\frac{2\pi}{3})$ and the second-order rotation $\rot(\ee_{1},\pi)$. It is of order $6$;
  \item $\DD_{4}$ is generated by the fourth-order rotation $\rot(\ee_{3},\frac{\pi}{2})$ and the second-order rotation $\rot(\ee_{1},\pi)$. It is of order $8$;
  \item $\octa$ is the \emph{octahedral} group, the orientation-preserving symmetry group of the cube with vertices $(\pm 1,\pm 1,\pm 1)$, which is of order 24;
  \item $\OO(2)$ is the group generated by all rotations $\rot(\ee_{3},\theta)$ ($\theta\in [0;2\pi[$) and the second-order rotation $\rot(\ee_{1},\pi)$. It is of infinite order.
\end{itemize}

In this list, the notation $\rot(\nn,\theta)$ denotes a rotation of angle $\theta$ around axis $\langle \nn \rangle$, with the convention that $\rot(\ee_{3},\theta)$ has the following matrix representation in the canonical basis $(\ee_{i})$  of $\RR^{3}$
\begin{equation*}
  \rot(\ee_{3},\theta)=
  \begin{pmatrix}
    \cos \theta & -\sin \theta & 0 \\
    \sin \theta & \cos \theta  & 0 \\
    0           & 0            & 1
  \end{pmatrix}.
\end{equation*}

\begin{figure}[h]\label{fig:lattice-ela}
  \centering
  \includegraphics[scale=1.3]{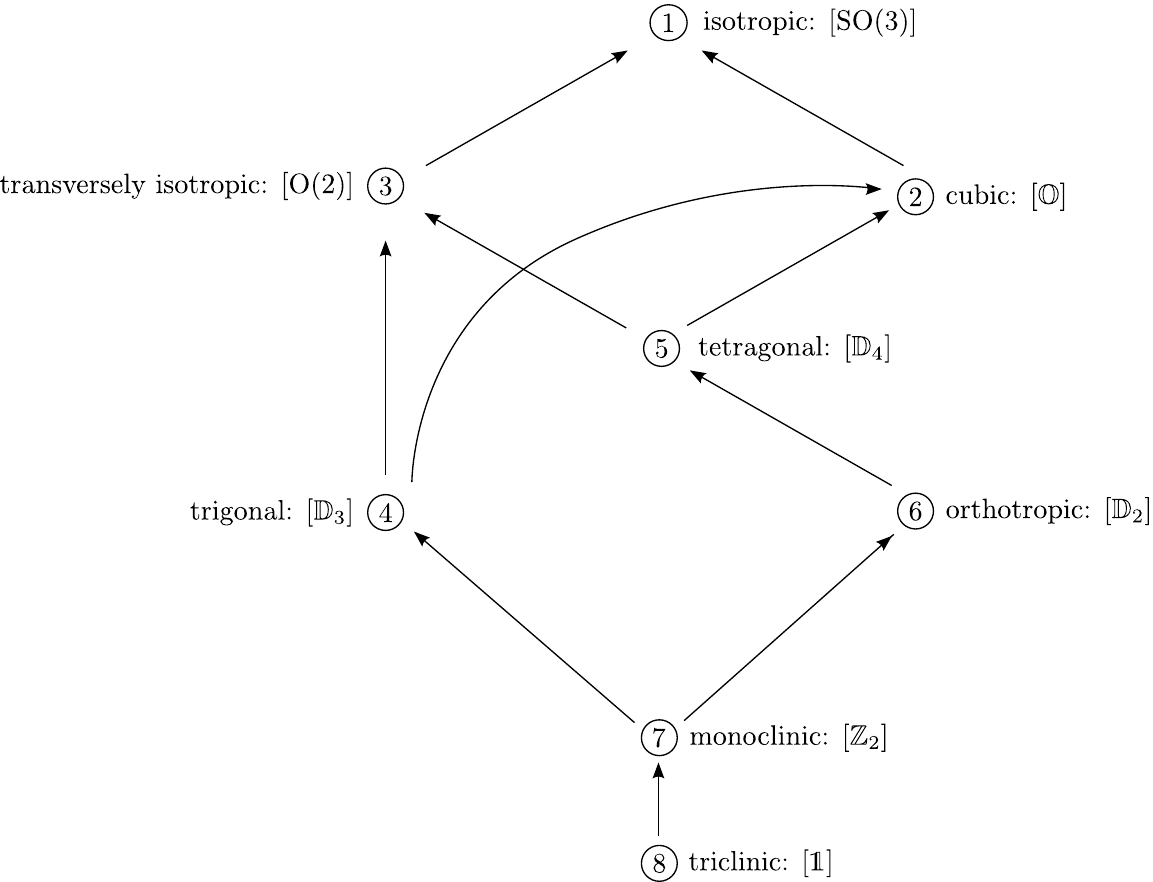}
  \caption{The eight symmetry classes of the Elasticity tensor~\cite{FV1996,AKP2014}.}
  \label{fig:lattice}
\end{figure}

There exists a \emph{partial order} on symmetry classes, induced by inclusion between subgroups, and defined as follows:
\begin{equation}
  [G_{1}] \preceq [G_{2}] \iff \exists g \in \SO(3), \quad G_{1} \subset g \, G_{2} \, g^{-1}.
\end{equation}
We can thus say that a tensor has ``at least'' or ``at most'' such or such symmetry. For example, a tensor $\bE$ is said to be at least orthotropic if it is either orthotropic, tetragonal, transversely-isotropic, cubic or isotropic. A tensor $\bE$ is said to be at least trigonal if it is either trigonal, transversely-isotropic, cubic or isotropic. This order is however \emph{partial}, which means that two classes cannot necessarily be compared (for example the trigonal and the tetragonal classes). The symmetry classes and their relations are summarized in Figure~\ref{fig:lattice}, where an arrow $[G_{1}]\rightarrow [G_{2}]$ means that $[G_{1}]\preceq [G_{2}]$.

For any subgroup $G$ of $\SO(3)$ in the list above and inducing a symmetry class $[G]$, we define its \emph{fixed point set}
\begin{equation*}
  \Ela^{G}: = \set{\bE\in \Ela; \; g\star \bE = \bE,\,\forall g\in G}.
\end{equation*}
This linear subspace of $\Ela$ is called a \emph{linear slice}. It meets all the orbits of tensors which have at least the symmetry class $[G]$. In other words, given an Elasticity tensor $\bE$ in the symmetry class $[G]$, there exists a rotation $g\in \SO(3)$ such that the symmetry group of $g\star \bE$ is exactly the subgroup $G$, which means that $g\star \bE \in \Ela^{G}$. We say then that the Elasticity tensor $g\star \bE$ is a \emph{normal form} of $\bE$.

\begin{rem}
  When $G$ is a finite group, the linear slice $\Ela^{G}$ is the subspace of solutions of the linear system $g_{k}\star \bE = \bE$ ($k = 1,\dotsc,r$), where $g_{k}$ generate $G$.
\end{rem}

We recall now, for each (non trivial) symmetry class $[G]$ of $\Ela$, a normal form for each class in Voigt's representation. An orthonormal basis in which Voigt's representation $[\bE]$ of an Elasticity tensor $\bE$ is a normal form is called a \emph{proper basis} or a \emph{natural basis} for $\bE$.

$\bullet$ The cubic normal form is defined as the subspace of tensors fixed by $\octa$. It has 3 independent parameters and writes
\begin{equation}\label{eq:NF-cubic}
  [\bE_{\octa}] =
  \begin{pmatrix}
    E_{1111} & E_{1112} & E_{1112} & 0        & 0        & 0        \\
    E_{1112} & E_{1111} & E_{1112} & 0        & 0        & 0        \\
    E_{1112} & E_{1112} & E_{1111} & 0        & 0        & 0        \\
    0        & 0        & 0        & E_{1212} & 0        & 0        \\
    0        & 0        & 0        & 0        & E_{1212} & 0        \\
    0        & 0        & 0        & 0        & 0        & E_{1212} \\
  \end{pmatrix}.
\end{equation}

$\bullet$ The transversely-isotropic normal form is defined as the subspace of tensors fixed by $\OO(2)$, for which the transverse isotropy axis is thus $\ee_{3}$. It has 5 independent parameters and writes
\begin{equation}\label{eq:NF-trans-iso}
  [\bE_{\OO(2)}] =
  \begin{pmatrix}
    E_{1111} & E_{1122} & E_{1133} & 0        & 0        & 0                                \\
    E_{1122} & E_{1111} & E_{1133} & 0        & 0        & 0                                \\
    E_{1133} & E_{1133} & E_{3333} & 0        & 0        & 0                                \\
    0        & 0        & 0        & E_{1313} & 0        & 0                                \\
    0        & 0        & 0        & 0        & E_{1313} & 0                                \\
    0        & 0        & 0        & 0        & 0        & \frac{1}{2} (E_{1111}-E_{1122} ) \\
  \end{pmatrix}.
\end{equation}

$\bullet$ The trigonal normal form is defined as the subspace of tensors fixed by $\DD_{3}$. It has 6 independent parameters and writes
\begin{equation}\label{eq:NF-trigo}
  [\bE_{\DD_{3}}] =
  \begin{pmatrix}
    E_{1111} & E_{1122}    & E_{1133} & E_{1123}   & 0        & 0                               \\
    E_{1122} & E_{1111}    & E_{1133} & - E_{1123} & 0        & 0                               \\
    E_{1133} & E_{1133}    & E_{3333} & 0          & 0        & 0                               \\
    E_{1123} & -  E_{1123} & 0        & E_{1313}   & 0        & 0                               \\
    0        & 0           & 0        & 0          & E_{1313} & E_{1123}                        \\
    0        & 0           & 0        & 0          & E_{1123} & \frac{1}{2} (E_{1111}-E_{1122}) \\
  \end{pmatrix}.
\end{equation}

$\bullet$ The tetragonal normal form is defined as the subspace of tensors fixed by $\DD_{4}$. It has 6 independent parameters and writes
\begin{equation}\label{eq:NF-tetra}
  [\bE_{\DD_{4}}] =
  \begin{pmatrix}
    E_{1111} & E_{1122} & E_{1133} & 0        & 0        & 0        \\
    E_{1122} & E_{1111} & E_{1133} & 0        & 0        & 0        \\
    E_{1133} & E_{1133} & E_{3333} & 0        & 0        & 0        \\
    0        & 0        & 0        & E_{1313} & 0        & 0        \\
    0        & 0        & 0        & 0        & E_{1313} & 0        \\
    0        & 0        & 0        & 0        & 0        & E_{1212}
  \end{pmatrix}.
\end{equation}

$\bullet$ The orthotropic normal form is defined as the subspace of tensors fixed by $\DD_{2}$. It has 9 independent parameters and writes
\begin{equation}\label{eq:NF-ortho}
  [\bE_{\DD_{2}}] =
  \begin{pmatrix}
    E_{1111} & E_{1122} & E_{1133} & 0        & 0        & 0        \\
    E_{1122} & E_{2222} & E_{2233} & 0        & 0        & 0        \\
    E_{1133} & E_{2233} & E_{3333} & 0        & 0        & 0        \\
    0        & 0        & 0        & E_{2323} & 0        & 0        \\
    0        & 0        & 0        & 0        & E_{1313} & 0        \\
    0        & 0        & 0        & 0        & 0        & E_{1212} \\
  \end{pmatrix}.
\end{equation}

$\bullet$ The monoclinic normal form is defined as the subspace of tensors fixed by $\ZZ_{2}$. It has a symmetry plane normal to $\ee_{3}$, has 13 independent parameters and writes
\begin{equation}\label{eq:NF-mono}
  [\bE_{\ZZ_{2}}] =
  \begin{pmatrix}
    E_{1111} & E_{1122} & E_{1133} & 0        & 0        & E_{1112} \\
    E_{1122} & E_{2222} & E_{2233} & 0        & 0        & E_{2212} \\
    E_{1133} & E_{2233} & E_{3333} & 0        & 0        & E_{3312} \\
    0        & 0        & 0        & E_{2323} & E_{1323} & 0        \\
    0        & 0        & 0        & E_{1323} & E_{1313} & 0        \\
    E_{1112} & E_{2212} & E_{3312} & 0        & 0        & E_{1212} \\
  \end{pmatrix}.
\end{equation}

\section{Covariants of the Elasticity tensor}
\label{sec:covariants}

A \emph{polynomial covariant} $\bC(\bE)$ of the Elasticity tensor $\bE$, is a tensor $\bC$ which is a polynomial function of $\bE$ and such that
\begin{equation*}
  g \star \bC(\bE) = \bC(g\star \bE), \qquad \forall g \in \SO(3).
\end{equation*}

Examples of covariants are the dilatation and Voigt's second-order covariants $\bd(\bE)$, $\bv(\bE)$, defined by~\eqref{eq:dv}. A fourth-order covariant $\bH$ appears in the \emph{harmonic decomposition}
\begin{equation*}
  \bE = (\tr \bd, \tr \bv, \bd^{\prime}, \bv^{\prime}, \bH)
\end{equation*}
of an Elasticity tensor (see~\cite{Bac1970,Ona1984}), where $\bd^{\prime}$ and $\bv^{\prime}$ are the \emph{deviatoric parts} of $\bd$ and $\bv$, defined as
\begin{equation*}
  (\cdot)': = (\cdot)-\frac{1}{3}\tr(\cdot)\Idd.
\end{equation*}
More precisely, we can write
\begin{equation}\label{eq:H}
  \bE = \bE_{iso} + \bE_{\bd \bv} + \bH ,
\end{equation}
where the isotropic part of $\bE$ is defined as
\begin{equation*}
  \bE_{iso} : = \frac{1}{9}(\tr \bd) \; \Idd \otimes \Idd  +  \frac{1}{15}(3\tr\bv- \tr\bd) \; \bJ, \qquad \bJ: = \Idd  \otimesbar \Idd-\frac{1}{3}\Idd  \otimes \Idd,
\end{equation*}
and its dilatation-Voigt part as
\begin{equation*}
  \bE_{\bd \bv}: = \frac{1}{7}\left(\Idd \otimes (5 \bd^{\prime} -4 \bv^{\prime})+ (5 \bd^{\prime} -4 \bv^{\prime})\otimes \Idd\right)
  + \frac{2}{7}\left(\Idd \otimesbar (3 \bv^{\prime}-2 \bd^{\prime}) +(3 \bv^{\prime}-2 \bd^{\prime}) \otimesbar \Idd\right).
\end{equation*}
The remaining part
\begin{equation*}
  \bH: = \bE- \bE_{\bd \bv}-\bE_{iso},
\end{equation*}
is a fourth-order harmonic tensor (\textit{i.e.} totally symmetric and traceless). The detailed expression of its components $H_{ijkl}$ can be found in~\cite{Cow1989,Bae1993}.

\begin{rem}
  In these formulas, we have used the tensor products $\otimes$ and $\otimesbar$ of two symmetric second-order tensors $\ba$ and $\bb$, defined as
  \begin{equation*}
    (\ba \otimes \bb)_{ijkl} = a_{ij}b_{kl},
    \qquad
    (\ba \otimesbar \bb)_{ijkl} = \frac{1}{2}(a_{ik}b_{jl}+a_{il}b_{jk}).
  \end{equation*}
\end{rem}

The covariants $\bd(\bE)$, $\bv(\bE)$ and $\bH(\bE)$ depend linearly on $\bE$ but there are other non linear covariants which are extremely useful to study the geometry of $\bE$ and which have been used in~\cite{OKDD2018b} to formulate simple characterizations of the Elasticity symmetry classes. One of them is the following second-order quadratic covariant, first introduced by Boehler and coworkers in~\cite{BKO1994}:
\begin{equation*}
  \bd_{2}: = \tr_{13}(\bH :\bH) = \bH\threedots\bH, \qquad (\bd_{2})_{ij} = H_{ipqr}H_{pqrj}.
\end{equation*}
It depends on $\bE$ through $\bH$ and one has
\begin{equation}\label{eq:trd2}
  \tr\bd_{2} = \norm{\bH}^2.
\end{equation}
A full set of $70$ polynomial covariants of $\bH$ which generates the polynomial covariant algebra of $\bH$ has been produced in~\cite{OKDD2018b}.

\section{Recovering normal forms using covariants}
\label{sec:recovering-normal-forms}

Covariants are useful to characterize the symmetry class of a tensor~\cite{OKDD2018b}. For instance, we have introduced in~\eqref{eq:cross-product}, the generalized cross product  which writes
\begin{equation*}
  (\ba\times \bb)_{ijk} = \frac{1}{6} \left(b_{ip}\varepsilon_{pjq}a_{qk} + b_{ip}\varepsilon_{pkq}a_{qj}+b_{jp}\varepsilon_{piq}a_{qk} + b_{jp}\varepsilon_{pkq}a_{qi} + b_{kp}\varepsilon_{piq}a_{qj}+ b_{kp}\varepsilon_{pjq}a_{qi} \right),
\end{equation*}
for two second-order symmetric tensors $\bA=\ba$ and $\bB=\bb$ (see~\eqref{eq:axbijk} in~\autoref{sec:detailed-components} for the detailed expression of each component), and we have the following result~\cite{OKDD2018b}.

\begin{lem}
  A second-order symmetric tensor $\ba$ is orthotropic if and only if the third order covariant $\ba^{2}\times \ba$ is non-vanishing.
\end{lem}

\begin{rem}
  $\ba\times \Idd = 0$, $\ba'\times \bb = \ba\times \bb' = \ba'\times\, \bb'$.
\end{rem}

Consider now a family of second-order symmetric tensors $\mathcal{F} = \set{\ba_{1},\ba_{2},\dotsc,\ba_{n}}$, with $n\geq 2$. Recall that the symmetry class $[G_{\mathcal{F}}]$ of $\mathcal{F}$ is the conjugacy class of the subgroup
\begin{equation*}
  G_{\mathcal{F}} = \bigcap_{i} G_{\ba_{i}} = \set{g\in \SO(3);\; g\star\ba_{i} = \ba_{i},\, \forall i},
\end{equation*}
and that such a family is either isotropic, transversely-isotropic, orthotropic, monoclinic or triclinic. We have moreover the following result~\cite{OKDD2018b}.

\begin{thm}\label{thm:n-quadratic-forms}
  Let $(\ba_{1}, \dotsc ,\ba_{n})$ be an $n$-tuple of second-order symmetric tensors. Then:
  \begin{enumerate}
    \item $(\ba_{1}, \dotsc ,\ba_{n})$ is \emph{isotropic} if and only if
          \begin{equation*}
            {\ba_{k}}^{\prime} = 0, \quad 1 \le k \le n ,
          \end{equation*}
          where ${\ba_{k}}^{\prime}$ is the deviatoric part of $\ba_{k}$.
    \item $(\ba_{1}, \dotsc ,\ba_{n})$ is \emph{transversely-isotropic} if and only if there exists $\ba_{j}$ such that
          \begin{equation*}
            {\ba_{j}}^{\prime} \ne 0, \qquad \ba_{j} \times \ba_{j}^{2} = 0,
          \end{equation*}
          and
          \begin{equation*}
            \ba_{j} \times \ba_{k} = 0, \quad 1 \le k \le n .
          \end{equation*}
    \item $(\ba_{1}, \dotsc ,\ba_{n})$ is \emph{orthotropic} if and only if
          \begin{equation*}
            \tr(\ba_{k} \times \ba_{l}) = 0, \quad 1 \le k,l \le n ,
          \end{equation*}
          and
          \begin{itemize}
            \item either there exists $\ba_{j}$ such that
                  \begin{equation*}
                    \ba_{j} \times \ba_{j}^{2} \ne 0 ;
                  \end{equation*}
            \item or there exists a pair $(\ba_{i},\ba_{j})$ such that
                  \begin{equation*}
                    \ba_{i} \times \ba_{j} \ne 0 .
                  \end{equation*}
          \end{itemize}
    \item $(\ba_{1}, \dotsc ,\ba_{n})$ is \emph{monoclinic} if and only if there exists a pair $(\ba_{i},\ba_{j})$ such that
          \begin{equation*}
            \pmb{\omega} := \tr(\ba_{i} \times \ba_{j}) \ne 0, \quad \text{and} \quad  (\ba_{k}\pmb{\omega}) \times \pmb{\omega} = 0, \quad 1 \le k \le n .
          \end{equation*}
  \end{enumerate}
\end{thm}

\begin{rem}
  If we define the commutator of $\ba_{i}$ and $\ba_{j}$ by $[\ba_{i}, \ba_{j}]=\ba_{i} \ba_{j}-\ba_{j} \ba_{i}$, then we have
  \begin{equation*}
    \tr(\ba_{i} \times \ba_{j}) = -\frac{1}{6} \bepsilon : [\ba_{i}, \ba_{j}].
  \end{equation*}
\end{rem}

\begin{figure}[h]
  \centering
  \includegraphics[scale=1]{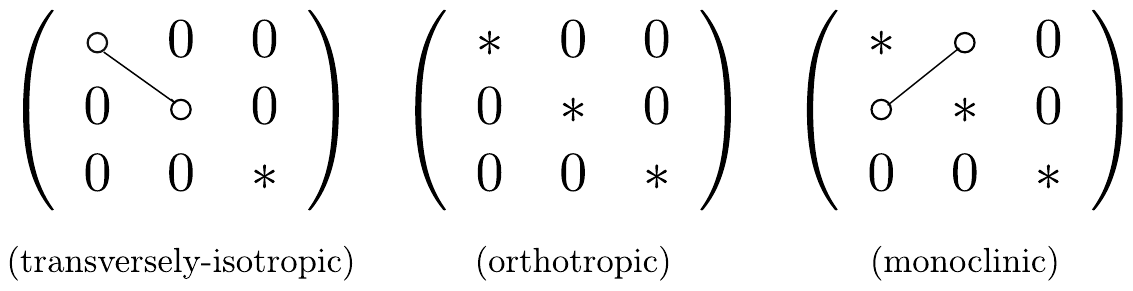}
  \caption{Shapes of normal forms for families of second-order symmetric tensors.}\label{fig:normal-forms}
\end{figure}

Theorem~\ref{thm:n-quadratic-forms} is the key point to recover the natural basis of a family $\mathcal{F} = \set{\ba_{1},\ba_{2},\dotsc,\ba_{n}}$ of second-order symmetric tensors as follows. A natural basis for the family $\mathcal{F}$ is one in which all the members of the family have the same matrix-shape with a maximum of zero (see~\autoref{fig:normal-forms}, where the joined circles mean ``equal components'' and the stars mean ``distinct components'').

The problem of finding a normal form for $\mathcal{F}$ is meaningful only when it is transversely-isotropic, orthotropic or monoclinic. As an illustration of our purpose, we shall now detail how to find a rotation which brings $\mathcal{F}$ into a normal form in each of these three cases.

\begin{description}
  \item[$\bullet$ $\mathcal{F}$ is transversely-isotropic] Find a member $\ba_{j}$ as in point (2) of theorem~\ref{thm:n-quadratic-forms}. Any basis in which $\ba_{j}$ is diagonal, the last vector corresponding to its simple eigenvalue, will achieve the task.
  \item[$\bullet$ $\mathcal{F}$ is orthotropic] If there exists $\ba_{j}$ orthotropic in $\mathcal{F}$ (\textit{i.e} ${\ba_{j}}^{2} \times \ba_{j} \ne 0$), then just diagonalize $\ba_{j}$ and this will answer the question. Otherwise, by point (3) of theorem~\ref{thm:n-quadratic-forms}, we can find an orthotropic pair $(\ba_{i},\ba_{j})$ in $\mathcal{F}$ where both $\ba_{i}$ and $\ba_{j}$ are transversely-isotropic. In that case, the eigenspaces of $\ba_{i}$ and $\ba_{j}$ corresponding to single eigenvalues are one-dimensional and mutually orthogonal. A natural basis for $\mathcal{F}$ is obtained by choosing a unit vector $\uu$ in the first space, $\vv$ in the second space and completing to a (direct) basis by adding $\uu \times \vv$.
  \item[$\bullet$ $\mathcal{F}$ is monoclinic] In that case, construct $\bomega$ as in point (4) of theorem~\ref{thm:n-quadratic-forms}. Normalize it to a unit vector and complete it into a basis by adding an orthonormal basis of the plane $\bomega^{\bot}$. Permute, if necessary, the vectors to obtain a direct basis and we are done.
\end{description}

\begin{rem}\label{rem:TI-simple-eigenvalue}
  If $\bt$ is a non-vanishing \emph{transversely-isotropic deviator}, we do not need to solve a polynomial equation to compute its unique simple eigenvalue. It is given by
  \begin{equation*}
    \lambda = 2 \dfrac{\tr(\bt^{ 3})}{\tr (\bt^{ 2})},
  \end{equation*}
  and the main axis of $\bt$ (eigenspace of the simple eigenvalue) corresponds to the one-dimensional subspace
  \begin{equation}\label{eq:simple-eigenvalue}
    \ker \left(\bt - 2 \dfrac{\tr(\bt^{ 3})}{\tr (\bt^{ 2})} \Idd\right).
  \end{equation}
\end{rem}

\begin{rem}\label{rem:orthotropic-couple}
  If $(\ba_{1},\ba_{2})$ is an orthotropic couple where both $\ba_{1}$ and $\ba_{2}$ are transversely-isotropic, then, their respective main axes are orthogonal and correspond respectively to
  \begin{equation*}
    \ker \left(\bt -2 \dfrac{\tr(\ba_{1}^{ 3})}{\tr (\ba_{1}^{ 2})} \Idd\right) \quad \text{and} \quad \ker \left(\bt -2 \dfrac{\tr(\ba_{2}^{ 3})}{\tr (\ba_{2}^{ 2})} \Idd\right).
  \end{equation*}
\end{rem}

The methodology developed above for a family $\mathcal{F}$ of second-order symmetric tensors will allow us to find a natural basis of all Elasticity tensors $\bE$, provided they are either transversely-isotropic, tetragonal, trigonal, orthotropic or monoclinic. The isotropic case is trivial and the triclinic case will not be considered in this paper (even if it also possible to define some kind of normal form for a triclinic tensor \cite{Fed1968,Nor1989}). The cubic case will be treated at the end of this section. To start with, we recall the following result which is a corollary of theorem~\cite[Theorem 10.2]{OKDD2018b}. It allows us to solve the problem when $\bE$ is either transversely-isotropic, tetragonal or trigonal (details will be provided in~\autoref{sec:effective-computations}).

\begin{lem}\label{lem:transversely-isotropic}
  Let $\bE$ be a transversely-isotropic, tetragonal or trigonal Elasticity tensor. Then, the triplet $(\bd^{\prime},\bv^{\prime},{\bd_{2}}^{\prime})$ is transversely-isotropic.
\end{lem}

To be able to reduce the case of an Elasticity tensor $\bE$ to a family of second-order symmetric tensors, when $\bE$ is either orthotropic or monoclinic, we need more second-order symmetric covariants which we shall introduce now. First, let us recall that the $2$-contraction between $\bH$ and a second-order tensor $\ba$ is defined as
\begin{equation*}
  (\bH:\ba)_{ij}: = H_{ijpq}a_{pq}.
\end{equation*}
Using this operation, we produce first the following two covariants
\begin{equation*}
  \bc_{3}: = \bH:\bd_{2}, \quad \text{and} \quad \bc_{4}: = \bH:\bc_{3},
\end{equation*}
and introduce two families of symmetric second-order covariants of $\bE$, which will allow us to solve the problem when $\bE$ is either orthotropic or monoclinic. The first family,
\begin{equation}\label{eq:orthotropic-family}
  \mathcal{F}_{o} : = \set{\bd^{\prime}, \bv^{\prime},{\bd_{2}}^{\prime},\bc_{3}, \bc_{4}, \bH : \bd, \bH : \bv, \bH : \bd^{2}, \bH : \bv^{2}},
\end{equation}
will be used in the orthotropic case and the second family,
\begin{equation}\label{eq:monoclinic-family}
  \mathcal{F}_{m} : = \left\{\bd^{\prime}, \bv^{\prime},{\bd_{2}}^{\prime}, \bc_{3}, \bc_{4}, \bH : \bd, \bH : \bv, \bH : \bd^{2}, \bH : \bv^{2},
  \bH : (\bd\bv)^{s}, \bH : (\bd\bd_{2})^{s}, \bH : (\bv\bd_{2})^{s}\right\},
\end{equation}
will be used in the monoclinic case. Here, $(\cdot)^s$ stands for the symmetric part of a second-order tensor. The key-point to conclude is the following result, which is a consequence of~\cite[Theorem 10.2]{OKDD2018b} and~\cite[Remark 10.4]{OKDD2018b}.

\begin{thm}\label{thm:orthotropic-monoclinic}
  For any Elasticity tensor $\bE$:
  \begin{enumerate}
    \item $\bE$ is monoclinic if and only if the family $\mathcal{F}_{m}$ of second-order tensors is monoclinic.
    \item $\bE$ is orthotropic if and only if the family $\mathcal{F}_{o}$ of second-order tensors is orthotropic.
    \item Moreover, if the family $\mathcal{F}_{o}$ is transversely isotropic, then, the triplet $(\bd^{\prime},\bv^{\prime},{\bd_{2}}^{\prime})$ is transversely isotropic and $\bE$ is either tetragonal, trigonal or transversely isotropic.
  \end{enumerate}
\end{thm}

It remains to solve the problem when $\bE$ is cubic. In that case, each second-order covariant of $\bE$ is isotropic~\cite{OKDD2018b}. Therefore its fourth-order covariant $\bH$ is necessarily cubic (and thus non-vanishing). A natural basis for $\bH$ is therefore also one for $\bE$. The key-point to calculate such a natural basis is then provided by the following theorem.

\begin{thm}\label{thm:cubic-harmonic-tensor}
  Let $\bH$ be a fourth-order cubic harmonic tensor. Then, the solutions of the linear equation
  \begin{equation}\label{eq:cubic-equation}
    \tr(\bH \times \ba) = 0,
  \end{equation}
  where $\ba$ is a second-order symmetric tensor, is a three-dimensional vector space. Moreover, orthotropic tensors $\ba$ which are solution of~\eqref{eq:cubic-equation} form a dense open set and the natural basis of any such orthotropic tensor is a natural basis for $\bH$.
\end{thm}

\begin{rem}
  This means that solving the linear system~\eqref{eq:cubic-equation} and picking-up randomly a solution among them provides us with an orthotropic second-order symmetric tensor $\ba$ which eigenvectors define a proper basis for $\bH$.
\end{rem}

\begin{rem}
  In an orthonormal basis, the 10 components of the totally symmetric three-order tensor $\tr(\bH \times \ba)$ write
  \begin{equation*}
    (\tr(\bH \times \ba))_{ijk} = \frac{1}{10} \left(\varepsilon_{ipq}H_{jkpr} + \varepsilon_{jpq}H_{ikpr} + \varepsilon_{kpq}H_{ijpr}\right)a_{qr}.
  \end{equation*}
  The detailed expressions of each component of $\tr(\bH \times \ba)$ are provided by~\eqref{eq:TrSxaijk} in~\autoref{sec:detailed-components}.
\end{rem}

\begin{proof}
  The binary operation $\tr( \bH \times \ba)$ being covariant, solutions $\ba$ of $\tr(\bH \times \ba) = 0$ write as $g \star \ba_{0}$, where $\ba_{0}$ are the solutions of $\tr(\bH_{0} \times \ba_{0}) = 0$, and where $\bH_{0}$ is the normal form of $\bH$. This normal form $\bH_{0}$ (see for instance~\cite{AKP2014}) writes, in Voigt's representation~\eqref{eq:Voigt}, as
  \begin{equation*}
    [\bH_{0}] = \delta \left(
    \begin{array}{cccccc}
        8  & -4 & -4 & 0  & 0  & 0  \\
        -4 & 8  & -4 & 0  & 0  & 0  \\
        -4 & -4 & 8  & 0  & 0  & 0  \\
        0  & 0  & 0  & -4 & 0  & 0  \\
        0  & 0  & 0  & 0  & -4 & 0  \\
        0  & 0  & 0  & 0  & 0  & -4
      \end{array}
    \right),
  \end{equation*}
  where $\delta\neq 0$. It can be checked that the space of solutions of the equation $\tr(\bH_{0} \times \ba_{0}) = 0$ corresponds exactly to the three-dimensional vector space of diagonal tensors, in which orthotropic tensors are a dense open set. Hence, any natural basis for $\bH_{0}$ (there are 24 such ones) is a natural basis for any solution $\ba_{0}$ of $\tr(\bH_{0} \times \ba_{0}) = 0$. Conversely, any natural basis of an \emph{orthotropic} solution $\ba_{0}$ corresponds to a natural basis of $\bH_{0}$, since there are only 24 such bases. Therefore, any natural basis for $\bH$ corresponds to a natural basis of $\ba$, an orthotropic solution of $\tr( \bH \times \ba)=0$, and \textit{vice-versa}, which ends the proof.
\end{proof}

\section{Experimental data}
\label{sec:experiment}

As pointed out in the introduction, natural bases have been obtained in the literature for some non-degenerate situations, using the dilatation or Voigt's tensors~\cite{CM1987,Cow1989,Jar1994,Bae1998,CVC2001}. We present here some data which will be used in~\autoref{sec:effective-computations} to illustrate our methodology. They consist in
\begin{itemize}
  \item a \emph{Ni base single crystal superalloy} (with its successive exactly cubic, tetragonal, orthotropic and monoclinic Elasticity approximations),
  \item an \emph{$\alpha$-quartz} (which belongs to the trigonal crystal system, its exactly trigonal Elasticity approximation~\cite{Sut1992,ZTL2013} is therefore considered),
  \item an exactly transversely-isotropic approximation issued from~\cite{KS2009}.
\end{itemize}

The \emph{Ni base single crystal superalloy} has a nearly cubic microstructure~\cite{FBG1998}. Thus, for this material, we will consider first the best cubic approximation of its associated Elasticity tensor obtained by François and coworkers by a minimization procedure fully described in \cite{Fra1995,FBG1996,FBG1998} (and for which the covariants are at least cubic, thus as second order deviatoric tensors $\bd^{\prime} = \bv^{\prime} = 0$). We will consider then successive relevant approximations, still obtained by a minimization procedure, preserving the symmetry planes of initial exactly cubic symmetry, and keeping $\bd^{\prime} = \bv^{\prime} = 0$ (in order to illustrate the abilities of proposed methodology to handle the degenerate cases). These approximations will exactly belong to one of the tetragonal, orthotropic and monoclinic symmetry classes.

Note that in this section and \autoref{sec:effective-computations}, all experimental tensors, as well as their linear covariants, are expressed in GPa and that their fourth-order harmonic components are provided in~\autoref{sec-app:harmonic-component}. Moreover, all Voigt's (matrix) representations are given in the same fixed orthonormal basis $(\ee_{1}, \ee_{2}, \ee_{3})$.

\subsection{Ni base single crystal superalloy}
\label{subsec:Ni-superalloy}

Voigt's representation of the measured Elasticity tensor $\bE^{\gamma}$ of a \emph{Ni base single crystal superalloy}, obtained by ultrasonic measurements in~\cite{FBG1998}, writes as
\begin{equation*}
  [\bE^{\gamma}] =
  \begin{pmatrix}
    243 & 136 & 135 & 22  & 52  & -17 \\
    136 & 239 & 137 & -28 & 11  & 16  \\
    135 & 137 & 233 & 29  & -49 & 3   \\
    22  & -28 & 29  & 133 & -10 & -4  \\
    52  & 11  & -49 & -10 & 119 & -2  \\
    -17 & 16  & 3   & -4  & -2  & 130
  \end{pmatrix}.
\end{equation*}
We get
\begin{equation*}
  \tr \bd = 1531,
  \quad
  \tr \bv = 1479,
\end{equation*}
and
\begin{equation*}
  \bd^{\prime} =
  \begin{pmatrix}
    \tfrac{11}{3} & 2            & 14             \\
    2             & \tfrac{5}{3} & 23             \\
    14            & 23           & -\tfrac{16}{3}
  \end{pmatrix}
  ,\qquad
  \bv^{\prime} =
  \begin{pmatrix}
    -1  & -11 & -1 \\
    -11 & 9   & -1 \\
    -1  & -1  & -8
  \end{pmatrix}.
\end{equation*}
The fourth-order harmonic part of $\bE^{\gamma}$ writes
\begin{equation*}
  [\bH] =
  \begin{pmatrix}
    -\scriptstyle\frac{1986}{35} & \scriptstyle\frac{1093}{35}  & \scriptstyle\frac{893}{35}   & 5                           & \scriptstyle\frac{352}{7}  &
    -\scriptstyle\frac{99}{7}                                                                                                                                                           \\
    \scriptstyle\frac{1093}{35}  & -\scriptstyle\frac{2306}{35} & \scriptstyle\frac{1213}{35}  & -31                         & \scriptstyle\frac{3}{7}    &
    \scriptstyle\frac{132}{7}                                                                                                                                                           \\
    \scriptstyle\frac{893}{35}   & \scriptstyle\frac{1213}{35}  & -\scriptstyle\frac{2106}{35} & 26                          & -\scriptstyle\frac{355}{7} &
    -\scriptstyle\frac{33}{7}                                                                                                                                                           \\
    5                            & -31                          & 26                           & \scriptstyle\frac{1213}{35} & -\scriptstyle\frac{33}{7}  & \scriptstyle\frac{3}{7}     \\
    \scriptstyle\frac{352}{7}    & \scriptstyle\frac{3}{7}      & -\scriptstyle\frac{355}{7}   & -\scriptstyle\frac{33}{7}   & \scriptstyle\frac{893}{35} & 5
    \\
    -\scriptstyle\frac{99}{7}    & \scriptstyle\frac{132}{7}    & -\scriptstyle\frac{33}{7}    & \scriptstyle\frac{3}{7}     & 5                          & \scriptstyle\frac{1093}{35}
  \end{pmatrix}.
\end{equation*}
We get thus
\begin{equation*}
  \norm{\bE^{\gamma}}^{2} = \norm{\bE_{iso}}^{2} + \norm{\bE_{\bd \bv}}^{2} + \norm{\bH}^{2} ,
\end{equation*}
where $\norm{\bE^{\gamma}}^{2} : = E^{\gamma}_{ijkl}E^{\gamma}_{ijkl}$. Moreover:
\begin{itemize}
  \item The squared norm of the isotropic part $\bE_{iso}$ of $\bE^{\gamma}$ writes
        \begin{equation*}
          \norm{\bE_{iso}}^{2} = \frac{1}{15} (2 (\tr \bd)^{2} - 2 \tr \bd \tr \bv + 3 (\tr \bv)^{2}),
        \end{equation*}
        and corresponds to the contribution of the isotropic parts of the dilatation and the Voigt tensors.
  \item The squared norm of the dilatation-Voigt parts $\bE_{\bd \bv}$ of $\bE^{\gamma}$ writes
        \begin{equation*}
          \norm{\bE_{\bd \bv}}^{2} = \frac{2}{21} \norm{\bd^{\prime}+2 \bv^{\prime}}^{2} +\frac{4}{3}\norm{\bd^{\prime}- \bv^{\prime}}^{2},
        \end{equation*}
        and corresponds to the contribution of the deviatoric parts of the dilatation and the Voigt tensors.
  \item The squared norm of the harmonic part $\bH$ of $\bE^{\gamma}$ writes
        \begin{equation}
          \norm{\bH}^{2} = \tr \bd_{2}.
        \end{equation}
\end{itemize}

For this single crystal superalloy, the isotropic contribution is the largest,
\begin{equation*}
  \frac{\norm{\bE_{iso}}^{2}}{\norm{\bE^{\gamma}}^{2}} = 0.880438 \, ,
\end{equation*}
while the anisotropic dilatational-Voigt contribution is negligible, as
\begin{equation*}
  \frac{\norm{\bE_{\bd \bv}}^{2}}{\norm{\bE^{\gamma}}^{2}} = 0.005826 \, ,
\end{equation*}
and the fourth-order harmonic contribution is of order two in magnitude,
\begin{equation*}
  \frac{\norm{\bH}^{2}}{\norm{\bE^{\gamma}}^{2}} = 0.113736 \, .
\end{equation*}

All the following approximations have identical isotropic parts.

\begin{description}

  \item[Approximation of $\bE^\gamma$ by a cubic Elasticity tensor]
        \begin{equation}\label{eq:cubic-experimental}
          [\bE^{\gamma}_{cubic}] =
          \begin{pmatrix}
            240.131  & 144.442  & 125.760  & 6.39666  & 41.9737  & -21.1614 \\
            144.442  & 223.957  & 141.935  & -27.7808 & 2.27754  & 16.6041  \\
            125.760  & 141.935  & 242.638  & 21.3841  & -44.2512 & 4.55736  \\
            6.39666  & -27.7808 & 21.3841  & 133.268  & 4.55736  & 2.27754  \\
            41.9737  & 2.27754  & -44.2512 & 4.55736  & 117.094  & 6.39666  \\
            -21.1614 & 16.6041  & 4.55736  & 2.27754  & 6.39666  & 135.776
          \end{pmatrix}.
        \end{equation}
        which is exactly cubic with $\norm{\bE^{\gamma}-\bE^{\gamma}_{cubic}}/\norm{\bE^{\gamma}} = 0.105$.

  \item[Approximation of $\bE^\gamma$ by a tetragonal Elasticity tensor]
        \begin{equation}\label{eq:ETetraRD}
          [\bE^{\gamma}_{tetra}] =
          \begin{pmatrix}
            239.333  & 141.383  & 129.618 & 5.8239   & 46.7414 & -21.0897 \\
            141.383  & 229.664  & 139.287 & -25.7103 & 1.8896  & 15.3674  \\
            129.618  & 139.287  & 241.429 & 19.8864  & -48.631 & 5.7223   \\
            5.8239   & -25.7103 & 19.8864 & 130.62   & 5.7223  & 1.8896   \\
            46.7414  & 1.8896   & -48.631 & 5.7223   & 120.951 & 5.8239   \\
            -21.0897 & 15.3674  & 5.7223  & 1.8896   & 5.8239  & 132.716
          \end{pmatrix},
        \end{equation}
        which is exactly tetragonal with $\norm{\bE^{\gamma}-\bE^{\gamma}_{tetra}}/\norm{\bE^{\gamma}} = 0.0996$.

  \item[First approximation of $\bE^\gamma$ by an orthotropic Elasticity tensor]
        \begin{equation}\label{eq:OrthExpTensor}
          [\bE^{\gamma\,(1)}_{orth}] =
          \begin{pmatrix}
            241.3079 & 139.4923 & 129.533  & 5.3342   & 46.3021  & -20.3543 \\
            139.4923 & 229.6671 & 141.1738 & -26.2801 & 2.8831   & 14.4327  \\
            129.533  & 141.1738 & 239.6264 & 20.9459  & -49.1853 & 5.9215   \\
            5.3342   & -26.2801 & 20.9459  & 132.5072 & 5.9215   & 2.8831   \\
            46.3021  & 2.8831   & -49.1853 & 5.9215   & 120.8663 & 5.3342   \\
            -20.3543 & 14.4327  & 5.9215   & 2.8831   & 5.3342   & 130.8256
          \end{pmatrix},
        \end{equation}
        which is exactly orthotropic with $\norm{\bE^{\gamma}-\bE^{\gamma\,(1)}_{orth}}/\norm{\bE^{\gamma}} = 0.0988$.

  \item[Second approximation of $\bE^\gamma$ by an orthotropic Elasticity tensor $\bE^{\gamma\,(2)}_{orth}$]
        \begin{equation}\label{eq:OrthExpTensorPrime}
          [\bE^{\gamma\,(2)}_{orth}] =
          \begin{pmatrix}
            237.183  & 141.391  & 128.075  & 5.73042  & 42.1382  & -21.4889 \\
            141.391  & 228.24   & 142.671  & -27.3681 & 3.32484  & 16.2765  \\
            128.075  & 142.671  & 241.302  & 21.7967  & -44.0867 & 4.55057  \\
            5.73042  & -27.3681 & 21.7967  & 133.637  & 4.397    & 1.83613  \\
            42.1382  & 3.32484  & -44.0867 & 4.397    & 114.865  & 6.93608  \\
            -21.4889 & 16.2765  & 4.55057  & 1.83613  & 6.93608  & 137.635
          \end{pmatrix},
        \end{equation}
        which is exactly orthotropic with $\norm{\bE^{\gamma}-\bE^{\gamma\,(2)}_{orth}}/\norm{\bE^{\gamma}} = 0.1029$.

  \item[Approximation of $\bE^\gamma$ by a monoclinic Elasticity tensor $\bE^{\gamma}_{mono}$]
        \begin{equation}\label{eq:monoclinic-experimental}
          [\bE^{\gamma}_{mono}] =
          \begin{pmatrix}
            240.532  & 140.501  & 129.3    & 0.6715   & 47.7714  & -18.1515 \\
            140.501  & 231.37   & 138.463  & -26.3969 & 4.4758   & 18.2628  \\
            129.3    & 138.463  & 242.57   & 25.7254  & -52.2472 & -0.1113  \\
            0.6715   & -26.3969 & 25.7254  & 129.796  & -0.1113  & 4.4758   \\
            47.7714  & 4.4758   & -52.2472 & -0.1113  & 120.634  & 0.6715   \\
            -18.1515 & 18.2628  & -0.1113  & 4.4758   & 0.6715   & 131.834
          \end{pmatrix},
        \end{equation}
        which is exactly monoclinic with $\norm{\bE^{\gamma}-\bE^{\gamma}_{mono}}/\norm{\bE^{\gamma}} = 0.0883$.

\end{description}

\subsection{Trigonal $\alpha$-quartz}
\label{subsec:alpha-quartz}

We consider now the trigonal Elasticity tensor of the \emph{$\alpha$-quartz}, provided in~\cite{ZTL2013} and obtained from experimental data issued from~\cite{Sut1992}. In Voigt's representation, it writes
\begin{equation}\label{eq:trigonal-experimental}
  [\bE^{\alpha}_{trig}] =
  \begin{pmatrix}
    7.9122  & 0.7161  & 2.1801  & -0.0778 & -0.054  & -0.6541 \\
    0.7161  & 10.808  & -0.0235 & 0.4322  & -0.0725 & 2.1674  \\
    2.1801  & -0.0235 & 10.2544 & 0.8267  & 0.5779  & -1.2035 \\
    -0.0778 & 0.4322  & 0.8267  & 4.3259  & -1.0971 & 0.0825  \\
    -0.054  & -0.0725 & 0.5779  & -1.0971 & 6.2917  & 0.3278  \\
    -0.6541 & 2.1674  & -1.2035 & 0.0825  & 0.3278  & 4.5151
  \end{pmatrix},
\end{equation}
which is exactly trigonal.

\subsection{Transversely isotropic Elasticity tensor}
\label{subsec:ti-Oak}

Finally, we will consider the transversely-isotropic Elasticity tensor $\bE^{KS}_{TI}$, obtained in~\cite{KS2009}. In Voigt's representation, it writes
\begin{equation}\label{eq:EIT}
  [\bE^{KS}_{TI}] =
  \begin{pmatrix}
    1.4373  & 0.5382  & 0.2699 & -0.0879 & -0.181  & -0.054  \\
    0.5382  & 1.4577  & 0.2634 & -0.1688 & -0.1009 & -0.0553 \\
    0.2699  & 0.2634  & 1.0327 & 0.0295  & 0.0324  & 0.0348  \\
    -0.0879 & -0.1688 & 0.0295 & 0.4046  & 0.0588  & -0.0209 \\
    -0.181  & -0.1009 & 0.0324 & 0.0588  & 0.4156  & -0.015  \\
    -0.054  & -0.0553 & 0.0348 & -0.0209 & -0.015  & 0.4615
  \end{pmatrix},
\end{equation}
which is exactly transversely isotropic and which was obtained as the closest transversely-isotropic tensor to the following raw Elasticity tensor
\begin{equation*}
  [\bE^{KS}] =
  \begin{pmatrix}
    1.3045  & 0.6327  & 0.2592 & -0.1039 & -0.2385 & -0.1215 \\
    0.6327  & 1.4131  & 0.2648 & -0.1261 & -0.0705 & -0.0301 \\
    0.2592  & 0.2648  & 1.0389 & 0.0395  & 0.045   & 0.0317  \\
    -0.1039 & -0.1261 & 0.0395 & 0.4794  & 0.019   & -0.0514 \\
    -0.2385 & -0.0705 & 0.045  & 0.019   & 0.3747  & -0.016  \\
    -0.1215 & -0.0301 & 0.0317 & -0.0514 & -0.016  & 0.5128
  \end{pmatrix},
\end{equation*}
with relative error $\norm{\bE^{KS}-\bE^{KS}_{TI}}/\norm{\bE^{KS}} = 0.1278$.

\section{Effective computations}
\label{sec:effective-computations}

For each symmetry class, we shall explain how to find a rotation which brings an Elasticity tensor $\bE$ whose components are given in an arbitrary orthonormal direct basis $\mathcal{B}_{0}= (\ee_{1},\ee_{2},\ee_{3})$, into its normal form. More precisely, we will compute an orthonormal basis $\mathcal{B} = (\uu_{1},\uu_{2},\uu_{3})$, and hence a rotation
\begin{equation}\label{eq:rot-change-basis}
  g =
  \begin{pmatrix}
    \uu_{1}\cdot \ee_{1} & \uu_{1}\cdot \ee_{2} & \uu_{1}\cdot \ee_{3} \\
    \uu_{2}\cdot \ee_{1} & \uu_{2}\cdot \ee_{2} & \uu_{2}\cdot \ee_{3} \\
    \uu_{3}\cdot \ee_{1} & \uu_{3}\cdot \ee_{2} & \uu_{3}\cdot \ee_{3}
  \end{pmatrix},
\end{equation}
such that the Voigt's representation~\eqref{eq:Voigt} of $g\star \bE$ is a normal form of the symmetry class of $\bE$. In the following, all Voigt's representations of Elasticity tensors are given in the initial orthonormal basis $(\ee_{1}, \ee_{2}, \ee_{3})$.

\subsection{Cubic class}
\label{subsec:cubic}

The proposed methodology for a cubic Elasticity tensor $\bE_{cubic}$ is the following.
\begin{enumerate}
  \item Calculate the fourth-order harmonic tensor $\bH$ of $\bE_{cubic}$ from~\eqref{eq:H}.
  \item Solve the linear system
        \begin{equation*}
          \tr( \bH \times \ba) = 0,
          \qquad\quad
          (\tr( \bH \times \ba))_{ijk}= \frac{1}{10}  \left(\varepsilon_{ipq}H_{jkpr} + \varepsilon_{jpq}H_{ikpr} + \varepsilon_{kpq}H_{ijpr}\right)a_{qr},
        \end{equation*}
        where $\ba$ is a second-order symmetric tensor and (see \autoref{sec:detailed-components} for the detailed components expressions for third order tensor $\tr(\bH\times \ba)$). Pick-up randomly a solution $\ba$ among them. According to theorem~\ref{thm:cubic-harmonic-tensor}, it will be orthotropic (\emph{i.e.} $\ba$ will have three distinct eigenvalues). This can be checked by verifying that $\ba^{2} \times \ba \ne0$.
  \item Diagonalize $\ba$ and compute a direct orthonormal basis $\mathcal{B} = (\uu_{1},\uu_{2},\uu_{3})$ of eigenvectors for $\ba$.
  \item The normal form \eqref{eq:NF-cubic} is given by $\bE_{\octa} = g\star \bE_{cubic}$ with $g$ defined by~\eqref{eq:rot-change-basis} and its action, by~\eqref{eq:gstarE}.
\end{enumerate}

\begin{rem}\label{rem:trHxaprime}
  Since $\tr( \bH \times \Idd) = 0$ for every tensor $\bH$, it is enough to solve the equation
  \begin{equation*}
    \tr( \bH \times \ba^{\prime}) = 0,
  \end{equation*}
  for deviatoric tensors $\ba^{\prime}$, which leads to solve a linear system in a five-dimensional space.
\end{rem}

\begin{ex}
  Consider the cubic Elasticity tensor~\eqref{eq:cubic-experimental} for Ni base single crystal superalloy. It is such that $\tr \bd = 1531$, $\tr \bv = 1479$, $\bd^{\prime} = 0$, $\bv^{\prime} = 0$.

  \begin{enumerate}
    \item Its fourth-order harmonic part $\bH$ is given by \eqref{eq:H-cubic-experimental} and we get ${\bd_{2}}^{\prime} = (\tr_{13}\bH^{2})' = 0$.

    \item  Setting arbitrarily $a_{13}' = 1$ and  $a_{12}' = 1$, the solution of $\tr( \bH \times \ba') = 0$ (see remark \ref{rem:trHxaprime}) leads to
          \begin{equation*}
            \ba' =
            \begin{pmatrix}
              3.485 & 1        & 1        \\
              1     & -9.93526 & -2.23089 \\
              1     & -2.23089 & 6.45025
            \end{pmatrix}
            ,
          \end{equation*}
          which is orthotropic since $(\ba')^{2} \times \ba'  \neq 0$.

    \item Computing a direct orthonormal basis of eigenvectors for $\ba'$, we get
          \begin{equation*}
            g =
            \begin{pmatrix}
              0.0813519 & -0.987342 & -0.136151 \\
              0.24438   & -0.112674 & 0.963111  \\
              -0.966261 & -0.111623 & 0.232121
            \end{pmatrix}.
          \end{equation*}

    \item And we can check that $(\bE^{\gamma}_{cubic})_\octa = g\star \bE^{\gamma}_{cubic}$ writes
          \begin{equation*}
            [(\bE^{\gamma}_{cubic})_\octa] =
            \begin{pmatrix}
              213.355 & 148.489 & 148.489 & 0       & 0       & 0       \\
              148.489 & 213.355 & 148.489 & 0       & 0       & 0       \\
              148.489 & 148.489 & 213.355 & 0       & 0       & 0       \\
              0       & 0       & 0       & 139.823 & 0       & 0       \\
              0       & 0       & 0       & 0       & 139.823 & 0       \\
              0       & 0       & 0       & 0       & 0       & 139.823
            \end{pmatrix}
            \quad \textrm{(GPa)}.
          \end{equation*}
          The normal form given in~\cite{FBG1998} is retrieved.
  \end{enumerate}
\end{ex}

\subsection{Transversely isotropic class}
\label{subsec:transversely-isotropic}

The proposed methodology for a transversely-isotropic Elasticity tensor $\bE_{TI}$ is the following.

\begin{enumerate}
  \item Compute the triplet of covariant deviators $(\bd^{\prime},\bv^{\prime},{\bd_{2}}^{\prime})$ of $\bE_{TI}$ (see~\autoref{sec:covariants}). By lemma~\ref{lem:transversely-isotropic}, the triplet $(\bd^{\prime},\bv^{\prime},{\bd_{2}}^{\prime})$ is transversely-isotropic. Thus, one of them, let us call it $\bt$, is transversely-isotropic.

  \item Let $\uu_{3}$ be the unit eigenvector corresponding to the single eigenvalue of $\bt$. By remark~\ref{rem:TI-simple-eigenvalue}, $\uu_{3}$ can be obtained by solving the linear system,
        \begin{equation*}
          \left(\bt -2 \dfrac{\tr(\bt^{ 3})}{\tr (\bt^{ 2})} \Idd\right)\uu = 0.
        \end{equation*}

  \item Complete $\uu_{3}$ into a direct orthonormal basis $\mathcal{B} = (\uu_{1},\uu_{2},\uu_{3})$ of $\RR^{3}$ by choosing an orthonormal pair $(\uu_{1},\uu_{2})$ orthogonal to $\uu_{3}$. For instance, if $\uu_{3}\neq \pm \ee_{3}$, one can choose
        \begin{equation}\label{eq:Ortho_Basis_Complement}
          \uu_{1} = \frac{\ee_{3}\times \uu_{3}}{\norm{\ee_{3}\times \uu_{3}}}
          \quad\textrm{and}\quad
          \uu_{2} = \uu_{3} \times \uu_{1}.
        \end{equation}
  \item The normal form \eqref{eq:NF-trans-iso} is given by $\bE_{\OO(2)} = g \star \bE_{TI}$ with rotation $g$ defined by~\eqref{eq:rot-change-basis} and its action on $\bE_{TI}$ computed using  \eqref{eq:gstarE}.
\end{enumerate}

\begin{rem}\label{rem:tIT}
  If $\bt=\bt'$ is a transversely-isotropic deviator, then $\bt^{2} \times \bt =0$ with $\bt \neq 0$, and $\bt$ writes $\bt=(-\frac{1}{2}\lambda, -\frac{1}{2}\lambda, \lambda)$ in its proper basis, where its simple eigenvalue $\lambda$ is derived as in remark~\ref{rem:TI-simple-eigenvalue}.
\end{rem}

\begin{ex}
  Consider the transversely-isotropic Elasticity tensor~\eqref{eq:EIT}.
  \begin{enumerate}
    \item We find
          \begin{equation*}
            \bd^{\prime} =
            \begin{pmatrix}
              0.221833 & -0.0745  & -0.2495   \\
              -0.0745  & 0.235733 & -0.2272   \\
              -0.2495  & -0.2272  & -0.457567
            \end{pmatrix},
          \end{equation*}
          \begin{equation*}
            \bv^{\prime} = 0.679222\, \bd^{\prime},
            \qquad
            {\bd_{2}}^{\prime} = -0.0977232 \, \bd^{\prime} ,
          \end{equation*}
          and we can check that the triplet $(\bd^{\prime},\bv^{\prime},{\bd_{2}}^{\prime})$ is transversely-isotropic. We choose $\bt = \bd^{\prime}$. Its simple eigenvalue is given by
          \begin{equation*}
            \lambda = 2\dfrac{\tr(\bd^{\prime 3})}{\tr(\bd^{\prime 2})} = -0.607173.
          \end{equation*}

    \item Solving the linear system~\eqref{eq:simple-eigenvalue} returns:
          \begin{equation*}
            \uu_{3} =
            \begin{pmatrix}
              0.29966  \\
              0.272898 \\
              0.914183
            \end{pmatrix}.
          \end{equation*}

    \item Build a direct orthonormal basis $\mathcal{B} = (\uu_{1},\uu_{2},\uu_{3})$ from~\eqref{eq:Ortho_Basis_Complement} and compute $g$ using~\eqref{eq:rot-change-basis}:
          \begin{equation*}
            g =
            \begin{pmatrix}
              -0.673321 & 0.73935   & 0        \\
              -0.675902 & -0.615539 & 0.405301 \\
              0.29966   & 0.272898  & 0.914183
            \end{pmatrix}.
          \end{equation*}

    \item  Then, we can check that $(\bE_{TI}^{KS})_{\OO(2)} = g \star \bE^{KS}_{TI}$ writes
          \begin{equation}
            [(\bE_{TI}^{KS})_{\OO(2)}] =
            \begin{pmatrix}
              1.5641 & 0.6046 & 0.1583 & 0      & 0      & 0      \\
              0.6046 & 1.5642 & 0.1582 & 0      & 0      & 0      \\
              0.1583 & 0.1582 & 1.0997 & 0      & 0      & 0      \\
              0      & 0      & 0      & 0.3258 & 0      & 0      \\
              0      & 0      & 0      & 0      & 0.3257 & 0      \\
              0      & 0      & 0      & 0      & 0      & 0.4799 \\
            \end{pmatrix}
            \quad \textrm{(GPa)},
          \end{equation}
          which is the normal form of the optimal transversely-isotropic tensor obtained in~\cite{KS2009}.
  \end{enumerate}
\end{ex}

\subsection{Trigonal class}
\label{subsec:trigonal}

The proposed methodology for a trigonal Elasticity tensor $\bE_{trig}$ is the following. The first three steps are the same as those for the transversely-isotropic case.

\begin{enumerate}
  \item Compute the transversely-isotropic triplet $( \bd^{\prime},\bv^{\prime},{\bd_{2}}^{\prime})$ from $\bE_{trig}$ (see lemma~\ref{lem:transversely-isotropic} and \autoref{sec:covariants}). Extract from this triplet a transversely-isotropic deviator $\bt$ (remark \ref{rem:tIT}).
  \item Let $\uu_{3}$ be a unit vector, solution of the linear system
        \begin{equation*}
          \left(\bt -2 \dfrac{\tr(\bt^{ 3})}{\tr (\bt^{ 2})} \Idd\right)\uu = 0.
        \end{equation*}
  \item Complete $\uu_{3}$ into a direct orthonormal basis $\mathcal{B}_{1} = (\ww_{1},\ww_{2},\uu_{3})$ of $\RR^{3}$, using~\eqref{eq:Ortho_Basis_Complement}, for instance, and define $g_{1}$ as the rotation given by~\eqref{eq:rot-change-basis}.

  \item Compute $\overline{\bE}: = g_{1}\star \bE_{trig}$ (using \eqref{eq:gstarE}) and define $\theta_{0}$ to be one solution of the equation
        \begin{equation}\label{eq:3theta}
          \overline{E}_{1123} \sin 3 \theta = \overline{E}_{1113} \cos 3 \theta.
        \end{equation}
  \item The normal form~\eqref{eq:NF-trigo} of $\bE_{trig}$ is given by $\bE_{\DD_{3}}=\rot(\ee_{3},\theta_{0})\star \overline{\bE}$, where $\rot(\ee_{3},\theta_{0})$ is the rotation of angle $\theta_{0}$ around axis $\ee_{3}$ and its action on $\overline{\bE}$ is computed using \eqref{eq:gstarE}.
\end{enumerate}

\begin{rem}
  Equation~\eqref{eq:3theta} derives from the observation that the matrix form of a trigonal Elasticity tensor with correct third axis $\uu_{3}$ writes
  \begin{equation}\label{eq:Ebar-trigo}
    [\overline\bE_{trig}] =
    \begin{pmatrix}
      \overline E_{1111} & \overline E_{1122}   & \overline E_{1133} & \overline E_{1123}   & \overline  E_{1113}  & 0                                                    \\
      \overline E_{1122} & \overline E_{1111}   & \overline E_{1133} & -\overline E_{1123}  & -\overline  E_{1113} & 0                                                    \\
      \overline E_{1133} & \overline E_{1133}   & \overline E_{3333} & 0                    & 0                    & 0                                                    \\
      \overline E_{1123} & - \overline E_{1123} & 0                  & \overline E_{1313}   & 0                    & -\overline  E_{1113}                                 \\
      \overline E_{1113} & - \overline E_{1113} & 0                  & 0                    & \overline E_{1313}   & \overline  E_{1123}                                  \\
      0                  & 0                    & 0                  & -\overline  E_{1113} & \overline  E_{1123}  & \frac{1}{2} (\overline E_{1111} -\overline E_{1122}) \\
    \end{pmatrix}.
  \end{equation}
  Thus, a rotation of $\overline\bE_{trig}$ around $\uu_{3}$ and of angle $\theta_{0}$, solution of \eqref{eq:3theta},
  leads to the normal form~\eqref{eq:NF-trigo}.
\end{rem}

\begin{ex}
  Consider the trigonal Elasticity tensor~\eqref{eq:trigonal-experimental} for $\alpha$-quartz.
  \begin{enumerate}
    \item  We compute
          \begin{equation*}
            \tr \bd = 34.709, \qquad \tr \bv = 59.249 \, ,
          \end{equation*}
          and
          \begin{equation*}
            \bv^{\prime}  =
            \begin{pmatrix}
              -1.02767 & 0.4162     & 0.6064  \\
              0.4162   & -0.0976667 & 1.5867  \\
              0.6064   & 1.5867     & 1.12533 \\
            \end{pmatrix}.
          \end{equation*}
          We check that $\bv^{\prime}$ is transversely-isotropic (${\bv^{\prime}}^{2}\times \bv^{\prime} = 0$) and observe that
          \begin{equation*}
            \bd^{\prime}     = 0.74434\, \bv^{\prime}, \quad \text{and} \quad {\bd_{2}}^{\prime} = -0.828279\, \bv^{\prime} .
          \end{equation*}
          The simple eigenvalue of $\bv^{\prime}$ is given by
          \begin{equation*}
            2\dfrac{\tr(\bv^{\prime 3})}{\tr(\bv^{\prime 2})} = 2.37334 \, .
          \end{equation*}

    \item Solving the linear system~\eqref{eq:simple-eigenvalue} with $\bt = \bv^{\prime}$ gives
          \begin{equation*}
            \uu_{3} =
            \begin{pmatrix}
              0.21137  \\
              0.553074 \\
              0.805873
            \end{pmatrix}
            ,\quad  \norm{\uu_{3}} = 1.
          \end{equation*}

    \item We build then a direct orthonormal basis $\mathcal{B}_{1} = (\ww_{1},\ww_{2},\uu_{3})$ using~\eqref{eq:Ortho_Basis_Complement} and define $g_{1}$ using~\eqref{eq:rot-change-basis},
          \begin{equation*}
            g_{1} =
            \begin{pmatrix}
              -0.934108 & 0.356991  & 0        \\
              -0.287689 & -0.752773 & 0.592088 \\
              0.21137   & 0.553074  & 0.805873
            \end{pmatrix}.
          \end{equation*}

    \item The Elasticity tensor $\overline{\bE} = g_{1} \star \bE_{trig}$ writes
          \begin{equation*}
            [\overline{\bE}] =
            \begin{pmatrix}
              8.76  & 0.6    & 1.33  & 1.706  & 0.289  & 0      \\
              0.6   & 8.76   & 1.33  & -1.706 & -0.288 & 0      \\
              1.33  & 1.33   & 10.68 & 0      & 0      & 0      \\
              1.706 & -1.706 & 0     & 5.72   & 0      & -0.288 \\
              0.289 & -0.288 & 0     & 0      & 5.72   & 1.706  \\
              0     & 0      & 0     & -0.288 & 1.706  & 4.08
            \end{pmatrix}.
          \end{equation*}

          We solve~\eqref{eq:3theta} and choose the solution
          \begin{equation*}
            \theta_{0} = \frac{1}{3} \arctan\left(\frac{\overline{E}_{1113}}{\overline{E}_{1123}}\right) = 0.0558614 \, .
          \end{equation*}

    \item We define
          \begin{equation*}
            \rot(\ee_{3},\theta_{0}) =
            \begin{pmatrix}
              0.99844   & -0.0558324 & 0 \\
              0.0558324 & 0.99844    & 0 \\
              0         & 0          & 1
            \end{pmatrix}.
          \end{equation*}
          Finally, one can check that $(\bE_{trig}^{\alpha})_{\DD_{3}} = \rot(\ee_{3},\theta_{0}) \star \overline{\bE}$ writes
          \begin{equation*}
            [(\bE_{trig}^{\alpha})_{\DD_{3}}] =
            \begin{pmatrix}
              8.76 & 0.6   & 1.33  & 1.73  & 0    & 0    \\
              0.6  & 8.76  & 1.33  & -1.73 & 0    & 0    \\
              1.33 & 1.33  & 10.68 & 0     & 0    & 0    \\
              1.73 & -1.73 & 0     & 5.72  & 0    & 0    \\
              0    & 0     & 0     & 0     & 5.72 & 1.73 \\
              0    & 0     & 0     & 0     & 1.73 & 4.08 \\
            \end{pmatrix}
            \quad \textrm{(GPa)}.
          \end{equation*}
  \end{enumerate}
\end{ex}

\subsection{Tetragonal class}
\label{sec:tetragonal}

The methodology for a tetragonal Elasticity tensor $\bE_{tetra}$ is similar to the one used for the trigonal case.

\begin{enumerate}
  \item Compute the transversely-isotropic triplet $(\bd^{\prime},\bv^{\prime},{\bd_{2}}^{\prime})$ from $\bE_{tetra}$ (see lemma~\ref{lem:transversely-isotropic} and \autoref{sec:covariants}). Extract from this triplet a transversely-isotropic deviator $\bt$ (remark \ref{rem:tIT}).
  \item Let $\uu_{3}$ with $\norm{\uu_{3}} = 1$ be a solution of the linear system
        \begin{equation*}
          \left(\bt -2 \dfrac{\tr(\bt^{ 3})}{\tr (\bt^{ 2})} \Idd\right)\uu = 0 \, .
        \end{equation*}
  \item Complete $\uu_{3}$ into a direct orthonormal basis $\mathcal{B}_{1} = (\ww_{1},\ww_{2},\uu_{3})$ of $\RR^{3}$, using~\eqref{eq:Ortho_Basis_Complement}, for instance and define $g_{1}$ as the rotation given by~\eqref{eq:rot-change-basis}.
  \item Compute $\overline{\bE} : = g_{1} \star \bE_{tetra}$ (using \eqref{eq:gstarE}) and define $\theta_{0}$ to be one solution of
        \begin{equation}\label{eq:4theta}
          4\overline E_{1112} \cos 4 \theta = \left( 2 \overline E_{1212}+ \overline E_{1122}- \overline E_{1111}\right)\sin 4 \theta ,
        \end{equation}
        which always exists if $\overline{\bE}$ is tetragonal.
  \item The normal form \eqref{eq:NF-tetra} is given by $\bE_{\DD_{4}}=\rot(\ee_{3},\theta_{0})\star \overline{\bE}$, where $\rot(\ee_{3},\theta_{0})$ is the rotation of angle $\theta_{0}$ around $\ee_{3}$ and its action on $\overline{\bE}$ is computed using \eqref{eq:gstarE}.
\end{enumerate}

\begin{rem}
  Equation~\eqref{eq:4theta} derives from the observation that the matrix-form of a tetragonal Elasticity tensor with correct third axis $\uu_{3}$ writes
  \begin{equation}\label{eq:Ebar-tetra}
    [\overline{\bE}_{tetra}] =
    \begin{pmatrix}
      \overline E_{1111}  & \overline E_{1122}  & \overline E_{1133} & 0                   & 0                  & \overline E_{1112}  \\
      \overline E_{1122}  & \overline E_{1111}  &
      \overline  E_{1133} & 0                   & 0                  & -\overline E_{1112}                                            \\
      \overline E_{1133}  & \overline E_{1133}  & \overline E_{3333} & 0                   & 0                  & 0                   \\
      0                   & 0                   & 0                  & \overline E_{1313}  & 0                  & 0                   \\
      0                   & 0                   & 0                  & 0                   & \overline E_{1313} & 0                   \\
      \overline E_{1112}  & -\overline E_{1112} & 0                  & 0                   & 0                  & \overline{E}_{1212}
    \end{pmatrix}.
  \end{equation}
  Thus, a rotation of $\overline{\bE}_{tetra}$ around $\uu_{3}$ and of angle $\theta_{0}$, solution of \eqref{eq:4theta},
  leads to the normal form~\eqref{eq:NF-tetra}.
\end{rem}

\begin{ex}
  Consider the tetragonal Elasticity tensor~\eqref{eq:ETetraRD} for Ni base single crystal superalloy.
  \begin{enumerate}
    \item We get $\bd^{\prime} = 0$, $\bv^{\prime} = 0$ and
          \begin{align*}
            {\bd_{2}}^{\prime} & =
            \begin{pmatrix}
              1389.87 & 341.696  & 47.1186  \\
              341.696 & -2729.03 & -571.863 \\
              47.1186 & -571.863 & 1339.17
            \end{pmatrix},
          \end{align*}
          which is transversely-isotropic ($({\bd_{2}}^{\prime})^{2} \times {\bd_{2}}^{\prime} = 0$). Its simple eigenvalue is given by
          \begin{equation*}
            2\dfrac{\tr(\bd_{2}^{\prime 3})}{\tr(\bd_{2}^{\prime 2})} = -2836.05 \, .
          \end{equation*}

    \item Solving the linear system~\eqref{eq:simple-eigenvalue} with $\bt = {\bd_{2}}^{\prime}$ gives
          \begin{equation*}
            \uu_{3} =
            \begin{pmatrix}
              0.0813519 \\
              -0.987342 \\
              -0.136151
            \end{pmatrix}
            , \quad \norm{\uu_{3}} = 1.
          \end{equation*}

    \item We build then a direct orthonormal basis $\mathcal{B}_{1} = (\ww_{1},\ww_{2},\uu_{3})$ using~\eqref{eq:Ortho_Basis_Complement} and define $g_{1}$ using~\eqref{eq:rot-change-basis},
          \begin{equation*}
            g_{1} =
            \begin{pmatrix}
              0.996623  & 0.0821166 & 0         \\
              0.0111802 & -0.135691 & 0.990688  \\
              0.0813519 & -0.987342 & -0.136151
            \end{pmatrix}.
          \end{equation*}

    \item We compute $\overline{\bE} : = g_{1} \star \bE_{tetra}$ and solve~\eqref{eq:4theta} and choose the solution
          \begin{equation*}
            \theta_{0} = \frac{1}{4} \arctan\left(\frac{4 \overline E_{1112}}{2 \overline E_{1212}+ \overline E_{1122}- \overline E_{1111}}\right) = 0.236501 \, .
          \end{equation*}

    \item Finally, we can check that $(\bE^{\gamma}_{tetra})_{\DD_{4}}=\rot(\ee_{3},\theta_{0})\star \overline{\bE}$ writes
          \begin{equation*}
            [(\bE^{\gamma}_{tetra})_{\DD_{4}}] =
            \begin{pmatrix}
              210.103 & 154.993 & 145.237 & 0       & 0       & 0       \\
              154.993 & 210.103 & 145.237 & 0       & 0       & 0       \\
              145.237 & 145.237 & 219.858 & 0       & 0       & 0       \\
              0       & 0       & 0       & 136.571 & 0       & 0       \\
              0       & 0       & 0       & 0       & 136.571 & 0       \\
              0       & 0       & 0       & 0       & 0       & 146.326
            \end{pmatrix}.
          \end{equation*}
  \end{enumerate}
\end{ex}

\subsection{Orthotropic class}
\label{subsec:orthotropic}

The methodology for an orthotropic Elasticity tensor $\bE_{ortho}$ is based on the deep investigation of the family $\mathcal{F}_{o}$ of second-order symmetric covariants given by~\eqref{eq:orthotropic-family}. This family is orthotropic by theorem~\ref{thm:orthotropic-monoclinic} and we have to distinguish between two cases.
\begin{enumerate}
  \item If there exists an orthotropic tensor $\ba$ in the family $\mathcal{F}_{o}$, then, a direct orthonormal basis of eigenvectors for $\ba$ is also a natural basis for $\mathcal{F}_{o}$.
  \item Otherwise, we can find an orthotropic couple $(\ba_{1},\ba_{2})$ in $\mathcal{F}_{o}$. In that case, both $\ba_{1}$ and $\ba_{2}$ are transversely-isotropic and their respective main axis are orthogonal. Let $\uu_{1}$ and $\uu_{2}$ be unit vectors spanning these axes (they can be obtained using remark~\ref{rem:orthotropic-couple}). Then, a natural basis for $\bE_{ortho}$ is $\mathcal{B} := (\uu_{1},\uu_{2},\uu_{1}\times \uu_{2})$.
\end{enumerate}
In both cases, the orthotropic normal form~\eqref{eq:NF-ortho} is recovered by $(\bE_{ortho})_{\DD_{2}} = g\star \bE_{ortho}$, where $g$ is defined by~\eqref{eq:rot-change-basis}
and its action on $ \bE_{ortho}$ is computed using \eqref{eq:gstarE}.

In~\cite[Theorem 10.2]{OKDD2018b}, it was shown that if $\bE_{ortho}$ is orthotropic, then the triplet $(\bd^{\prime},\bv^{\prime},{\bd_{2}}^{\prime})$ is either orthotropic or transversely-isotropic. This observation leads to a possible optimization of the methodology proposed above.

\begin{itemize}
  \item If this triplet is orthotropic, our methodology can be optimized, by looking for an orthotropic tensor or an orthotropic couple of transversely-isotropic tensors in this triplet, rather than in the whole family $\mathcal F_{o}$.
  \item If the triplet is transversely-isotropic, an alternative methodology similar to the one used for a trigonal or a tetragonal tensor is still possible and is detailed below.
        \begin{enumerate}
          \item Extract a transversely-isotropic deviator $\bt$ from the triplet $(\bd^{\prime},\bv^{\prime},{\bd_{2}}^{\prime})$.
          \item Compute $\uu_{3}$ with $\norm{\uu_{3}} = 1$ as a solution of the linear system
                \begin{equation*}
                  \left(\bt -2 \dfrac{\tr(\bt^{ 3})}{\tr (\bt^{ 2})} \Idd\right)\uu = 0,
                \end{equation*}
                as explained in remark~\ref{rem:TI-simple-eigenvalue}.
          \item Complete $\uu_{3}$ into a direct orthonormal basis $\mathcal{B}_{1} = (\ww_{1},\ww_{2},\uu_{3})$ of $\RR^{3}$, using~\eqref{eq:Ortho_Basis_Complement} for instance, and define $g_{1}$ as the rotation given by~\eqref{eq:rot-change-basis}.

          \item Compute $\overline{\bE} : = g_{1} \star \bE_{ortho}$ (using \eqref{eq:gstarE}) and let $\theta_{0}$ be a solution of
                \begin{equation}\label{eq:ortho-theta}
                  2 \overline E_{3312} \cos 2 \theta = ( \overline E_{1133}-\overline E_{2233}) \sin 2 \theta ,
                \end{equation}
                which always exists as $\overline{\bE}$ is an orthotropic tensor.
          \item The normal form~\eqref{eq:NF-ortho} is given by $(\bE_{ortho})_{\DD_{2}} = \rot(\ee_{3},\theta_{0})\star \overline{\bE}$, where $\rot(\ee_{3},\theta_{0})$ is the rotation of angle $\theta_{0}$ around $\ee_{3}$ and its action on $\overline{\bE}$ is computed using \eqref{eq:gstarE}.
        \end{enumerate}
\end{itemize}

\begin{rem}
  As in the trigonal and the tetragonal cases, equation~\eqref{eq:ortho-theta} is derived from the observation that an orthotropic Elasticity tensor with one correct axis, say $\uu_{3}$, writes
  \begin{equation*}
    [\overline\bE_{ortho}] =
    \begin{pmatrix}
      \overline E_{1111} & \overline E_{1122} & \overline E_{1133} & 0                  & 0                  & \overline E_{1112} \\
      \overline E_{1122} & \overline E_{2222} & \overline E_{2233} & 0                  & 0                  & \overline E_{2212} \\
      \overline E_{1133} & \overline E_{2233} & \overline E_{3333} & 0                  & 0                  & \overline E_{3312} \\
      0                  & 0                  & 0                  & \overline E_{2323} & 0                  & 0                  \\
      0                  & 0                  & 0                  & 0                  & \overline E_{1313} & 0                  \\
      \overline E_{1112} & \overline E_{2212} & \overline E_{3312} & 0                  & 0                  & \overline E_{1212}
    \end{pmatrix}.
  \end{equation*}
  Thus, a rotation of $\overline\bE_{ortho}$ around $\uu_{3}$ and of angle $\theta_{0}$, solution of~\eqref{eq:ortho-theta}, leads to the normal form~\eqref{eq:NF-ortho}.
\end{rem}

\begin{ex}
  Consider the orthotropic Elasticity tensor~\eqref{eq:OrthExpTensor} for Ni base single crystal superalloy. This example is interesting because both the dilatation and the Voigt second-order covariants of this Elasticity tensor are isotropic, $\bd^{\prime} = \bv^{\prime} = 0$. Hence, simple methods to recover its normal form fail. However, one can check that its deviatoric second-order covariant
  \begin{equation*}
    {\bd_{2}}^{\prime} =
    \begin{pmatrix}
      523.33  & 207.103  & 500.816  \\
      207.103 & -2721.59 & -651.919 \\
      500.816 & -651.919 & 2198.26
    \end{pmatrix},
  \end{equation*}
  is orthotropic. Its diagonalization defines using~\eqref{eq:rot-change-basis} the rotation
  \begin{equation*}
    g =
    \begin{pmatrix}
      0.0813478 & -0.987343 & -0.136151 \\
      0.244376  & -0.112676 & 0.963112  \\
      -0.966262 & -0.111619 & 0.232117
    \end{pmatrix},
  \end{equation*}
  and one can check that the Elasticity tensor $(\bE^{\gamma}_{ortho})_{\DD_{2}} = g\star \bE^{\gamma\,(1)}_{orth}$ writes
  \begin{equation*}
    [(\bE^{\gamma}_{ortho})_{\DD_{2}}] =
    \begin{pmatrix}
      219.858 & 147.607 & 142.867 & 0       & 0     & 0      \\
      147.607 & 207.732 & 154.992 & 0       & 0     & 0      \\
      142.867 & 154.992 & 212.473 & 0       & 0     & 0      \\
      0       & 0       & 0       & 146.326 & 0     & 0      \\
      0       & 0       & 0       & 0       & 134.2 & 0      \\
      0       & 0       & 0       & 0       & 0     & 138.94
    \end{pmatrix}
    \qquad \text{(GPa)},
  \end{equation*}
  and corresponds to its orthotropic normal form~\eqref{eq:NF-ortho}.
\end{ex}

\begin{ex}
  Consider now the second orthotropic Elasticity tensor approximation~\eqref{eq:OrthExpTensorPrime} for Ni base single crystal superalloy. This time ${\bd_{2}}^{\prime} = 0$ (since its fourth-order harmonic part is cubic) and both $\bd^{\prime}$ and $\bv^{\prime}$ are transversely-isotropic but not of the same axis. The pair $(\bd^{\prime}, \bv^{\prime})$ is orthotropic. The unit eigenvectors $\uu_{1}$ and $\uu_{2}$ corresponding respectively to the simple eigenvalue of $\bd^{\prime}$ and $\bv^{\prime}$ are
  \begin{equation*}
    \uu_{1} =
    \begin{pmatrix}
      -0.966261 \\
      -0.111623 \\
      0.232121
    \end{pmatrix},
    \qquad
    \uu_{2} =
    \begin{pmatrix}
      0.0813519 \\
      -0.987342 \\
      -0.136151
    \end{pmatrix}.
  \end{equation*}

  The rotation $g$ build from \eqref{eq:rot-change-basis}, with $\uu_{3} = \uu_{1}\times \uu_{2}$, is such that the Elasticity tensor $(\bE^{\gamma\, \prime}_{ortho})_{\DD_{2}} = g\star \bE^{\gamma\,(1)}_{orth}$ has the orthotropic normal form
  \begin{equation*}
    [(\bE^{\gamma\, \prime}_{ortho})_{\DD_{2}}] =
    \begin{pmatrix}
      217.806 & 149.478 & 145.095 & 0.      & 0.      & 0.      \\
      149.478 & 212.006 & 150.896 & 0.      & 0.      & 0.      \\
      145.095 & 150.896 & 210.252 & 0.      & 0.      & 0.      \\
      0.      & 0.      & 0.      & 137.507 & 0.      & 0.      \\
      0.      & 0.      & 0.      & 0.      & 141.857 & 0.      \\
      0.      & 0.      & 0.      & 0.      & 0.      & 140.104
    \end{pmatrix}
    \qquad \text{(GPa)}.
  \end{equation*}
\end{ex}

\begin{rem}
  Compared to previous works~\cite{CM1987,Cow1989,Jar1994,Bae1998,CVC2001}, our procedure, relying on theorem~\ref{thm:orthotropic-monoclinic}, is exhaustive and allows to handle all degenerate cases. It is based on the list $\mathcal F_{o}$ of second-order covariants which carries all the information required to recover the normal form of an orthotropic Elasticity tensor.
\end{rem}

\subsection{Monoclinic class}
\label{subsec:monoclinic}

The methodology for a monoclinic Elasticity tensor $\bE_{mono}$ is based on the investigation of the family $\mathcal{F}_{m}$ of second-order symmetric covariants given by~\eqref{eq:monoclinic-family}. This family is monoclinic by theorem~\ref{thm:orthotropic-monoclinic}. The algorithm is the following.

\begin{enumerate}
  \item Find a common eigenvector $\bomega$ for all second-order covariants in the family $\mathcal{F}_{m}$, by computing the commutators ($\bomega = \bepsilon : [\ba_{i},\ba_{j}]$), as in theorem~\ref{thm:n-quadratic-forms}.
  \item Set $\uu_{3} = \pmb\omega/\norm{\pmb\omega}$ and complete it into a direct orthonormal basis $\mathcal{B} = (\uu_{1},\uu_{2},\uu_{3})$, using~\eqref{eq:Ortho_Basis_Complement} for instance.
  \item The monoclinic normal form~\eqref{eq:NF-mono} is given by $g\star \bE_{mono}$ where $g$ is defined by~\eqref{eq:rot-change-basis} and its action on $\bE_{mono}$ is computed using \eqref{eq:gstarE}.
\end{enumerate}

\begin{rem}
  In most (non degenerate) cases, the commutator $\bepsilon : [\bd, \bv]=2\,\bepsilon : (\bd \bv)$ of the dilatation and the Voigt tensors will allow to initiate the first step of the algorithm (as in~\cite{CM1987,Cow1989,Jar1994}). But $\bepsilon : [\bd, \bv]$ may vanish, as in the next example. In that case, another candidate is required (for instance $\bomega = \pmb\varepsilon:(\bd_{2} \bc_{3})$ in the next example). Our methodology relies on Theorem \ref{thm:orthotropic-monoclinic} and is exhaustive.
\end{rem}

\begin{ex}
  Consider the degenerate monoclinic Elasticity tensor~\eqref{eq:monoclinic-experimental}, where $\bd^{\prime} = \bv^{\prime} = 0$.
  \begin{enumerate}
    \item A non-vanishing first-order covariant is $\bomega = \bepsilon : (\bd_{2} \bc_{3})$ which writes
          \begin{equation*}
            \bomega = 10^7
            \begin{pmatrix}
              16.727   \\
              -7.71214 \\
              65.9218
            \end{pmatrix}.
          \end{equation*}

    \item Set
          \begin{equation*}
            \uu_{3} = \frac{1}{\norm{\bomega}}\bomega, \qquad  \uu_{1} = \dfrac{1}{\sqrt{n_{1}^{2}+n_{2}^{2}}}
            \begin{pmatrix}
              -n_{2} \\
              n_{1}  \\
              0
            \end{pmatrix}
            , \qquad  \uu_{2} = \uu_{3} \times \uu_{1} .
          \end{equation*}
          We get then
          \begin{equation*}
            \uu_{3} =
            \begin{pmatrix}
              0.24438   \\
              -0.112674 \\
              0.963111
            \end{pmatrix}
            ,\qquad \uu_{1} =
            \begin{pmatrix}
              0.418699 \\
              0.908125 \\
              0
            \end{pmatrix}
            , \qquad \uu_{2} =
            \begin{pmatrix}
              -0.874625 \\
              0.403254  \\
              0.269104
            \end{pmatrix},
          \end{equation*}
          and
          \begin{equation*}
            g =
            \begin{pmatrix}
              0.4187   & 0.90812  & 0       \\
              -0.87463 & 0.40325  & 0.2691  \\
              0.24438  & -0.11267 & 0.96311
            \end{pmatrix}.
          \end{equation*}

    \item One can check that $(\bE^{\gamma}_{mono})_{\ZZ_{2}}= g \star \bE_{mono}$ writes
          \begin{equation*}
            [(\bE^{\gamma}_{mono})_{\ZZ_{2}}] =
            \begin{pmatrix}
              299.7 & 68.6  & 142.1 & 0     & 0     & -42. \\
              68.6  & 281.3 & 160.5 & 0     & 0     & 41.1 \\
              142.1 & 160.5 & 207.7 & 0     & 0     & 0.9  \\
              0     & 0     & 0     & 151.8 & 0.9   & 0    \\
              0     & 0     & 0     & 0.9   & 133.4 & 0    \\
              -42.  & 41.1  & 0.9   & 0     & 0     & 59.9
            \end{pmatrix}
            \quad\textrm{(GPa)},
          \end{equation*}
          which is the normal form~\eqref{eq:NF-mono}.
  \end{enumerate}
\end{ex}

\begin{rem}
  Recall that, following \cite{Lek1963,Fed1968}, an additional zero can be placed in the normal form $\widehat{\bE}=\bE_{\ZZ_{2}}$ of a monoclinic tensor. This is due to the fact that any rotation around the third axis $\uu_{3}=\ee_{3}$ of the normal form~\eqref{eq:NF-mono} does not changes the shape of this normal form. If either
  \begin{equation*}
    \widehat{E}_{2233} - \widehat{E}_{1133} \quad \text{or} \quad \widehat{E}_{2323} - \widehat{E}_{1313}
  \end{equation*}
  does not vanish, then, we can look for a rotation $\rot(\ee_{3},\theta_{36})\star \widehat{\bE}$, where
  \begin{equation*}
    \tan (2\theta_{36}) = {2 \widehat E_{3312}}/{(\widehat E_{2233} - \widehat E_{1133})} ,
  \end{equation*}
  so that the component $(\rot(\ee_{3},\theta_{36})\star \widehat{\bE})_{3312}$, in row 3, column 6 and row 6, column 3
  of the new normal form~\eqref{eq:NF-mono} vanishes. Or, we can look for a rotation $\rot(\ee_{3},\theta_{45})\star \widehat{\bE}$, where
  \begin{equation*}
    \tan (2\theta_{45}) = {2 \widehat E_{1323}}/{(\widehat E_{2323} - \widehat E_{1313})} ,
  \end{equation*}
  so that the component $(\rot(\ee_{3},\theta_{45})\star \widehat{\bE})_{1323}$ in row 4, column 5 and row 5, column 4 of the new normal form~\eqref{eq:NF-mono} vanishes.
\end{rem}

\section{An algorithm to detect the symmetry class and recover a normal form}
\label{sec:algo}

In this final section, we formulate an algorithm, based on the previous calculations, which allows to obtain the exact symmetry class of any Elasticity tensor and its normal form. To achieve this task, we will use the two families of \emph{deviatoric second order covariants} of $\bE$ already introduced in~\autoref{sec:covariants},
\begin{equation*}
  \mathcal{F}_{o} : = \set{\bd^{\prime}, \bv^{\prime},{\bd_{2}}^{\prime},\bc_{3}, \bc_{4}, \bH : \bd, \bH : \bv, \bH : \bd^{2}, \bH : \bv^{2}},
\end{equation*}
and
\begin{equation*}
  \mathcal{F}_{m} : =\mathcal{F}_{o}\cup  \left\{\bH : (\bd\bv)^{s}, \bH : (\bd\bd_{2})^{s}, \bH : (\bv\bd_{2})^{s}\right\},
\end{equation*}
where $\bc_{3}=\bH:\bd_{2}$ and $ \bc_{4}=\bH:\bc_{3}$. The families $\mathcal{F}_{o}$ and $\mathcal{F}_m$ are used in the orthotropic and monoclinic cases. For an Elasticity tensor which is either transversely isotropic, trigonal or tetragonal, all its second order covariants are transversely isotropic with the same axis, say $\uu_{3}$. To determine this axis, and find a rotation which brings it to $\ee_{3}$, a common procedure was described in \autoref{subsec:transversely-isotropic}, \autoref{subsec:trigonal} and \autoref{sec:tetragonal}. It is summarized below as \emph{Procedure~\ref{proc:Basis_Triplet_ISOT}}.

\begin{proc}\label{proc:Basis_Triplet_ISOT}
  \textbf{Input} : A transversely isotropic, trigonal or tetragonal Elasticity tensor $\bE$, with components $E_{ijkl}$ in a fixed orthonormal basis $(\ee_i)$.

  \textbf{Output} : A rotation $g_{1}\in \SO(3)$ and an Elasticity tensor $\overline{\bE}=g_{1}\star \bE$ such that all its second order covariants are transversely isotropic and of axis $\langle \ee_{3} \rangle$, with components $\overline E_{ijkl}$ in the basis $(\ee_i)$.
  \begin{enumerate}
    \item Compute the triplet of covariant deviators $(\bd^{\prime},\bv^{\prime},{\bd_{2}}^{\prime})$ of $\bE$. One of them is transversely isotropic, call it $\bt$.

    \item Let $\uu_{3}=\nn$ be a unit eigenvector corresponding to the single eigenvalue of $\bt$, obtained by solving the linear system $\big(\bt -2 \frac{\tr(\bt^{ 3})}{\tr (\bt^{ 2})} \Idd\big)\uu = 0$.

    \item Complete $\uu_{3}$ into a direct orthonormal basis $\mathcal{B} = (\uu_{1},\uu_{2},\uu_{3})$ of $\RR^{3}$ by choosing an orthonormal pair $(\uu_{1},\uu_{2})$ orthogonal to $\uu_{3}$.

    \item Define $g_{1}$ as
          \begin{equation*}
            g_{1} =
            \begin{pmatrix}
              \uu_{1}\cdot \ee_{1} & \uu_{1}\cdot \ee_{2} & \uu_{1}\cdot \ee_{3} \\
              \uu_{2}\cdot \ee_{1} & \uu_{2}\cdot \ee_{2} & \uu_{2}\cdot \ee_{3} \\
              \uu_{3}\cdot \ee_{1} & \uu_{3}\cdot \ee_{2} & \uu_{3}\cdot \ee_{3}
            \end{pmatrix}
            \quad\textrm{ and then} \quad
            \overline{\bE}:=g_{1}\star \bE.
          \end{equation*}
  \end{enumerate}
\end{proc}

\begin{rem}\label{rem:P_cases}
  The output $\overline{\bE}$ of \emph{Procedure \ref{proc:Basis_Triplet_ISOT}} is necessary either transversely isotropic, trigonal or tetragonal with transverse isotropy axis $\langle \ee_{3}\rangle$.
  If the following conditions,
  \begin{equation*}
    \overline E_{1212}=\frac{1}{2}(\overline E_{1111}-\overline E_{1122}), \quad \text{and} \quad \overline E_{1112}= \overline E_{1123}=\overline E_{1113}=0,
  \end{equation*}
  hold, then, one gets the transversely isotropic normal form \eqref{eq:NF-trans-iso}. Otherwise, there is a rotation $\rot(\ee_{3}, \theta)$ such that $\overline{\bE}=\rot(\ee_{3}, \theta)\star \bE_{0}$, where $\bE_{0}$ is either a trigonal Elasticity tensor $\bE_{\DD_{3}}$ with Voigt's representation, the normal form~\eqref{eq:NF-trigo} or a tetragonal Elasticity tensor $\bE_{\DD_{4}}$ with Voigt's representation, the normal form~\eqref{eq:NF-tetra}. When $\bE_{0}=\bE_{\DD_{3}}$ is trigonal (with Voigt's representation~\eqref{eq:NF-trigo} and with components $E_{ijkl}$ in the basis $(\ee_i)$), the calculation of $\overline{\bE}=\rot(\ee_{3}, \theta)\star \bE_{\DD_{3}}$ (using \eqref{eq:gstarE}) leads to
  \begin{equation}
    \begin{cases}
      \overline  E_{1123} & =E_{1123} (2 \cos 2 \theta -1) \cos \theta
      \\
      \overline E_{1113}  & =-E_{1123}  (2 \cos 2 \theta +1)\sin \theta
    \end{cases},
  \end{equation}
  so that $\overline{E}_{1123}\neq 0$ or $\overline{E}_{1113}\neq 0$. If this is not the case, the considered Elasticity tensor is neither transversely isotropic, nor trigonal but tetragonal.
\end{rem}

\begin{figure}[h!]
  \begin{center}
    \begin{tikzpicture}[scale=0.85,line cap=round,line join=round,>=triangle 45,every node/.style={scale=0.85}]
      \node[draw,circle] (root) at (-1,0) {\begin{tabular}{c} Familly \\ $\mathcal{F}_{o}$ \end{tabular}};
      \node[draw,rectangle,fill=black!20] (Cas11) at (3,-5.5) {\small{$\mathcal{F}_{o}$ isotropic}};
      \node[right] (Cas111) at (14.3,-7) {$\mathbf{E}$ isotropic};
      \node[right] (Cas112) at (14,-5) {\begin{tabular}{l}
          $\mathbf{E}$ cubic \\
          \small{(\autoref{subsec:cubic})}
        \end{tabular}};
      \node[draw,rectangle,fill=black!20] (Cas12) at (3,-3) {\small{\begin{tabular}{c} $\mathcal{F}_{o}$ transversely \\ isotropic \end{tabular}}};
      \node[draw,rectangle] (Cas121) at (3,-1) {\small{\begin{tabular}{l}
            Compute $\overline{\mathbf{E}}$ \\
            using Procedure~\ref{proc:Basis_Triplet_ISOT}
          \end{tabular}}};
      \node[right] (Cas1211) at (14,-3.2) {\small{\begin{tabular}{l}
            $\mathbf{E}$ transversely isotropic \\
            (\autoref{subsec:transversely-isotropic})
          \end{tabular}}};
      \node[right] (Cas12121) at (14,-1) {\begin{tabular}{l}
          $\mathbf{E}$ trigonal \\
          \small{(\autoref{subsec:trigonal})}
        \end{tabular}};
      \node[right] (Cas1212) at (7,0.5) {$\small{\bullet}$};
      \node[right] (Cas12122) at (14,1.5) {\begin{tabular}{l}
          $\mathbf{E}$ tetragonal \\
          \small{(\autoref{sec:tetragonal})}
        \end{tabular}};
      \node[draw,rectangle,fill=black!20] (Cas13) at (2,3.5) {\small{$\mathcal{F}_{o}$ orthotropic}};
      \node[right] (Cas131) at (14,3.5) {\begin{tabular}{l}
          $\mathbf{E}$ orthotropic \\
          \small{(\autoref{subsec:orthotropic})}
        \end{tabular}};
      \node[draw,rectangle,fill=black!20] (Cas14) at (2,5.5) {\small{$\mathcal{F}_{o}$ monoclinic}};
      \node[draw,circle] (Cas141) at (5.5,5.5) {\begin{tabular}{c} Familly \\ $\mathcal{F}_m$ \end{tabular}};
      \node[right] (Cas1411) at (14,5.5) {\begin{tabular}{l}
          $\mathbf{E}$ monoclinic \\
          \small{(\autoref{subsec:monoclinic})}
        \end{tabular}};
      \node[draw,rectangle,fill=black!20] (Cas15) at (2,7.5) {\small{$\mathcal{F}_{o}$ triclinic}};
      \node[right] (Cas151) at (14.3,7.5) {{$\mathbf{E}$ triclinic}};
      \draw[->,>=stealth] (root) edge [out=-90,in=-180] (Cas11.west);
      \draw[->,>=stealth] (root) edge [out=-75,in=-180] (Cas12.west);
      \draw[->,>=stealth] (root) edge [out=45,in=-90] (Cas13.south);
      \draw[->,>=stealth] (root) edge [out=90,in=-180] (Cas14.west);
      \draw[->,>=stealth] (root)  edge [out=120,in=180] (Cas15.west);
      \draw[->,>=stealth] (Cas11) -- (Cas111.west) node[fill=black!10,midway,rectangle,draw]{\small{$\mathbf{d}_{2}=0$}} ;
      \draw[->,>=stealth] (Cas11) -- (Cas112.west) node[fill=black!10,midway,rectangle,draw]{\small{\begin{tabular}{l} $\mathbf{d}'_{2}=0$ \\$\mathbf{d}_{2}\neq 0$ \end{tabular} }};
      \draw[->,>=stealth] (Cas12) -- (Cas121);
      \draw[->,>=stealth] (Cas121.south east) -- (Cas1211.west) node[fill=black!10,midway,rectangle,draw]{\small{
          $\begin{cases}
              \overline E_{1112}  =\overline E_{1123}=\overline E_{1113}=0 \\
              \overline{E}_{1212} =\frac{1}{2}(\overline{E}_{1111}-\overline{E}_{1122})
            \end{cases}$}};
      \draw[->,>=stealth] (Cas121.north east) -- (Cas1212.west) node[fill=black!10,midway,rectangle,draw]{\small{Else}};
      \draw[->,>=stealth] (Cas1212) -- (Cas12121.west) node[fill=black!10,midway,rectangle,draw]{\small{$\overline{E}_{1123}\neq 0$ or $\overline{E}_{1113}\neq 0$}};
      \draw[->,>=stealth] (Cas1212) -- (Cas12122.west) node[fill=black!10,midway,rectangle,draw]{\small{$\overline{E}_{1123}=\overline{E}_{1113}=0$}};
      \draw[->,>=stealth] (Cas13) -- (Cas131.west);
      \draw[->,>=stealth] (Cas14) -- (Cas141.west);
      \draw[->,>=stealth] (Cas141) -- (Cas1411.west) node[fill=black!10,midway,rectangle,draw]{\small{$\mathcal{F}_m$ monoclinic}};
      \draw[->,>=stealth] (Cas141) -- (Cas151.south west) node[fill=black!10,midway,rectangle,draw]{\small{$\mathcal{F}_m$ triclinic}};
      \draw[->,>=stealth] (Cas15) -- (Cas151.west);
    \end{tikzpicture}
  \end{center}
  \caption{An algorithm to detect the symmetry class of an Elasticity tensor and recover its normal form.}
  \label{fig:Alg_Sym_Class}
\end{figure}
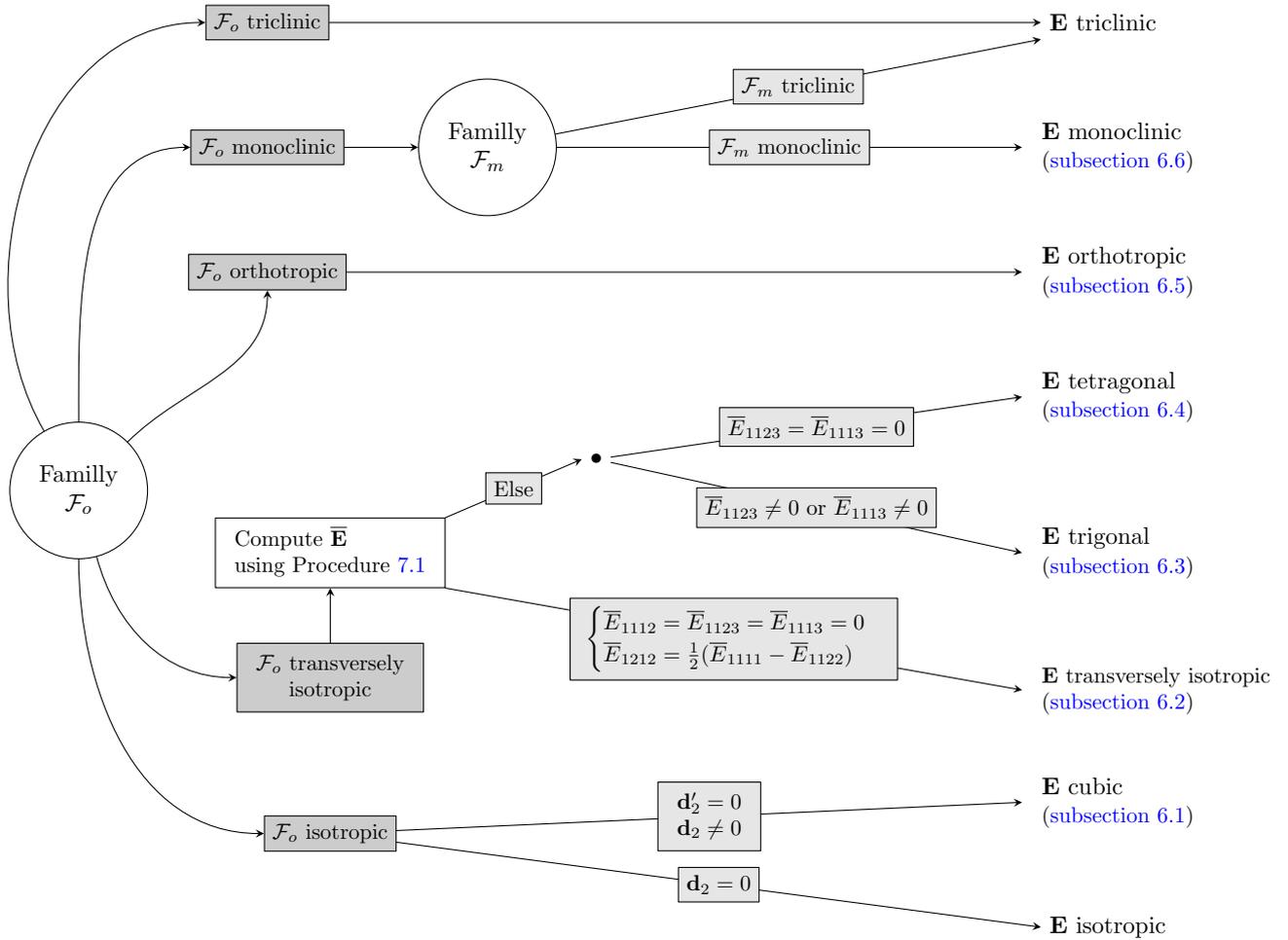

Finally, the algorithm to detect the symmetry class of a given Elasticity tensor is summarized in \autoref{fig:Alg_Sym_Class} and detailed below.
\begin{enumerate}
  \item If either the family $\mathcal{F}_{o}$ or $\mathcal{F}_m$ is triclinic, then $\bE$ is triclinic (else its covariants would have inherited its symmetry group \cite{BKO1994,OKDD2018b}).

  \item If both families $\mathcal{F}_{o}$ and $\mathcal{F}_m$ are monoclinic, then $\bE$ is monoclinic by theorem \ref{thm:orthotropic-monoclinic}(1).

  \item If $\mathcal{F}_{o}$ is orthotropic, then $\bE$ is orthotropic by theorem  \ref{thm:orthotropic-monoclinic}(2).

  \item If $\mathcal{F}_{o}$ is transversely isotropic, $\bE$ is then either tetragonal, trigonal or transversely isotropic by theorem  \ref{thm:orthotropic-monoclinic}(3). Using procedure~\ref{proc:Basis_Triplet_ISOT}, we build a new tensor $\overline{\bE}$ with components $\overline E_{ijkl}$ and by remark~\ref{rem:P_cases}:
        \begin{enumerate}
          \item if $\overline E_{1112}= \overline E_{1123}=\overline E_{1113}=0$ and $\overline{E}_{1212}=\frac{1}{2}(\overline{E}_{1111}-\overline{E}_{1122})$ then $\bE$ is transversely isotropic;
          \item otherwise if $\overline{E}_{1123}=\overline{E}_{1113}=0$ then $\bE$ is tetragonal, and if $\overline{E}_{1123}\neq 0$ or $\overline{E}_{1113}\neq 0$ then $\bE$ is trigonal.
        \end{enumerate}

  \item If $\mathcal{F}_{o}$ is isotropic, then, $\bd_{2}'=\bd'=\bv'=0$, thus $\bd_{2}= \frac{1}{3} \left(\tr \bd_{2}\right) \Idd$ and $\tr \bd_{2} = \norm{\bH}^2$ by~\eqref{eq:trd2}). In that case,
        \begin{enumerate}
          \item either $\bd_{2}=0$, so that $\bH=0$, and therefore $\bE=\bE_{iso}$ is isotropic by~\eqref{eq:H};
          \item or $\bd_{2}\neq 0$ and $\bH$ is cubic by~\cite[Theorem 9.3]{OKDD2018b} and so is $\bE$, by~\cite[Theorem 10.2]{OKDD2018b}.
        \end{enumerate}
\end{enumerate}

\section*{Conclusion}

We have formulated effective methods to recover the normal form of an Elasticity tensor, measured in any basis, provided that we know to which symmetry class it belongs to (this other problem having been solved, by the way, in a previous work~\cite{OKDD2018b}). For each symmetry class, a simple algorithm has been provided. In particular, thanks to the generalized cross product~\eqref{eq:cross-product} between totally symmetric tensors, a very simple method has been proposed to recover the normal form of an Elasticity tensor with cubic symmetry. It requires only to solve a linear system in five variables and diagonalize a three-dimensional symmetric matrix.

These procedures are moreover \emph{exhaustive}. All degenerate cases are handled, in particular, when second-order covariants, like $\bd^{\prime}$ and $\bv^{\prime}$, or first-order covariants like $\bepsilon : [\bd, \bv]$ vanish. To formulate and prove these results, we have used families of covariants derived in~\cite{OKDD2018b}, which are crucial to establish necessary and sufficient conditions for an Elasticity tensor to belong to a given symmetry class. As a by-product, applying our procedures to a given Elasticity tensor $\bE$ allows, not only to recognize a normal form for it, but is also a way to determine its symmetry class. An algorithm to achieve this task has been formulated in~\autoref{sec:algo}.

Besides, to illustrate our methods, we have applied them on experimental Elasticity tensors found in the literature, and this for each symmetry class.

Noises and experimental errors have not been considered since this subject is a full work in itself and will be addressed in forthcoming papers.

\appendix

\section{Harmonic components of considered Elasticity tensors}
\label{sec-app:harmonic-component}

In this section, all the linear covariants $\bd$, $\bv$, $\bH$ are given in GPa and the fourth-order harmonic part $\bH$ is expressed in Voigt's representation.

\begin{description}
  \item[Cubic approximation $\bE^\gamma_{cubic}$~\eqref{eq:cubic-experimental} of $\bE^\gamma$]
        $\,$ \newline
        $\bd^{\prime} = \bv^{\prime} = 0$, ${\bd_{2}}^{\prime} = 0$, $\tr \bd = 1531$, $\tr \bv = 1479$, and
        \begin{equation}\label{eq:H-cubic-experimental}
          [\bH^{\gamma}_{cubic}] =
          \begin{pmatrix}
            -59.1358 & 38.9089  & 20.2269  & 6.39666  & 41.9737  & -21.1614 \\
            38.9089  & -75.3102 & 36.4013  & -27.7808 & 2.27754  & 16.6041  \\
            20.2269  & 36.4013  & -56.6282 & 21.3841  & -44.2512 & 4.55736  \\
            6.39666  & -27.7808 & 21.3841  & 36.4013  & 4.55736  & 2.27754  \\
            41.9737  & 2.27754  & -44.2512 & 4.55736  & 20.2269  & 6.39666  \\
            -21.1614 & 16.6041  & 4.55736  & 2.27754  & 6.39666  & 38.9089
          \end{pmatrix}.
        \end{equation}

  \item[Tetragonal approximation $\bE^\gamma_{tetra}$~\eqref{eq:ETetraRD} of $\bE^\gamma$]
        $\,$ \newline
        $\bd^{\prime} = \bv^{\prime} = 0$, $\tr \bd = 1531$, $\tr \bv = 1479$, and
        \begin{equation}\label{eq:H-tetragonal-experimental}
          [\bH^{\gamma}_{tetra}] =
          \begin{pmatrix}
            -59.9342 & 35.8495  & 24.0847  & 5.8239   & 46.7414 & -21.0897 \\
            35.8495  & -69.6028 & 33.7533  & -25.7103 & 1.8896  & 15.3674  \\
            24.0847  & 33.7533  & -57.8381 & 19.8864  & -48.631 & 5.7223   \\
            5.8239   & -25.7103 & 19.8864  & 33.7533  & 5.7223  & 1.8896   \\
            46.7414  & 1.8896   & -48.631  & 5.7223   & 24.0847 & 5.8239   \\
            -21.0897 & 15.3674  & 5.7223   & 1.8896   & 5.8239  & 35.8495
          \end{pmatrix}.
        \end{equation}

  \item[First orthotropic approximation $\bE^{\gamma\, (1)}_{ortho}$~\eqref{eq:OrthExpTensor} of $\bE^\gamma$]
        $\,$ \newline
        $\bd^{\prime} = \bv^{\prime} = 0$, $\tr \bd = 1531$, $\tr \bv = 1479$ and
        \begin{equation}\label{eq:H-orthotropic-experimental}
          [\bH^{\gamma}_{orth}] =
          \begin{pmatrix}
            -57.9586 & 33.959   & 23.9997  & 5.3342   & 46.3021  & -20.3543 \\
            33.959   & -69.5995 & 35.6405  & -26.2801 & 2.88311  & 14.4327  \\
            23.9997  & 35.6405  & -59.6402 & 20.9459  & -49.1853 & 5.92151  \\
            5.3342   & -26.2801 & 20.9459  & 35.6405  & 5.92151  & 2.88311  \\
            46.3021  & 2.88311  & -49.1853 & 5.92151  & 23.9997  & 5.3342   \\
            -20.3543 & 14.4327  & 5.92151  & 2.88311  & 5.3342   & 33.959
          \end{pmatrix}.
        \end{equation}

  \item[Second orthotropic approximation $\bE^{\gamma\, (2)}_{ortho}$~\eqref{eq:OrthExpTensorPrime} of $\bE^\gamma$]
        $\,$ \newline
        $\tr \bd = 1531$, $\tr \bv = 1479$,
        \begin{equation*}
          \bd^{\prime} =
          \begin{pmatrix}
            -3.6837   & -0.661831 & 1.37627  \\
            -0.661831 & 1.96893   & 0.158989 \\
            1.37627   & 0.158989  & 1.71477
          \end{pmatrix}
          ,\, \quad
          \bv^{\prime}=
          \begin{pmatrix}
            -3.31669  & -0.8154 & -0.112441 \\
            -0.8154   & 6.51238 & 1.36466   \\
            -0.112441 & 1.36466 & -3.19569
          \end{pmatrix},
        \end{equation*}
        and $\bH = \bH^{\gamma}_{cubic}$ is given by~\eqref{eq:H-cubic-experimental} (in particular ${\bd_{2}}^{\prime} = 0$).

  \item[Monoclinic approximation $\bE^{\gamma}_{mono}$~\eqref{eq:monoclinic-experimental} of $\bE^\gamma$]
        $\,$ \newline
        $\bd^{\prime} = \bv^{\prime} = 0$, $\tr \bd = 1531$, $\tr \bv = 1479$ and
        \begin{equation}\label{eq:H-monoclinic-experimental}
          [\bH^{\gamma}_{mono}] =
          \begin{pmatrix}
            -58.7344 & 34.9674  & 23.767   & 0.6715   & 47.7714  & -18.1515 \\
            34.9674  & -67.8968 & 32.9294  & -26.3969 & 4.4758   & 18.2628  \\
            23.767   & 32.9294  & -56.6964 & 25.7254  & -52.2472 & -0.1113  \\
            0.6715   & -26.3969 & 25.7254  & 32.9294  & -0.1113  & 4.4758   \\
            47.7714  & 4.4758   & -52.2472 & -0.1113  & 23.767   & 0.6715   \\
            -18.1515 & 18.2628  & -0.1113  & 4.4758   & 0.6715   & 34.9674
          \end{pmatrix}.
        \end{equation}

  \item[Trigonal approximation $\bE^{\alpha}_{trig}$~\eqref{eq:trigonal-experimental} of $\alpha$-quartz Elasticity tensor]
        $\,$ \newline
        $\tr \bd = 34.72$, $\tr \bv = 59.24$,
        \begin{equation*}
          \bd^{\prime} =
          \begin{pmatrix}
            -0.764933 & 0.3098     & 0.4514   \\
            0.3098    & -0.0727333 & 1.1811   \\
            0.4514    & 1.1811     & 0.837667
          \end{pmatrix}
          ,\quad
          \bv^{\prime} =
          \begin{pmatrix}
            -1.02767 & 0.4162     & 0.6064  \\
            0.4162   & -0.0976667 & 1.5867  \\
            0.6064   & 1.5867     & 1.12533
          \end{pmatrix},
        \end{equation*}
        and
        \begin{equation}\label{eq:H-trigonal-experimental}
          [\bH^{\alpha}_{trig}] =
          \begin{pmatrix}
            -1.4953 & -0.0086 & 1.504   & -0.0148 & -0.2917 & -0.8173 \\
            -0.0086 & 0.6713  & -0.6626 & -0.1899 & -0.0484 & 2.0042  \\
            1.504   & -0.6626 & -0.8413 & 0.2046  & 0.3402  & -1.187  \\
            -0.0148 & -0.1899 & 0.2046  & -0.6626 & -1.187  & -0.0484 \\
            -0.2917 & -0.0484 & 0.3402  & -1.187  & 1.504   & -0.0148 \\
            -0.8173 & 2.0042  & -1.187  & -0.0484 & -0.0148 & -0.0086 \\
          \end{pmatrix}.
        \end{equation}

  \item[Transversely isotropic approximation $\bE^{KS}_{TI}$~\eqref{eq:EIT} of $\bE^{KS}$]
        $\,$ \newline
        $\tr\bd = 6.0707$, $\tr \bv = 6.4911$,
        \begin{equation*}
          \bd^{\prime} =
          \begin{pmatrix}
            0.221833 & -0.0745  & -0.2495   \\
            -0.0745  & 0.235733 & -0.2272   \\
            -0.2495  & -0.2272  & -0.457567
          \end{pmatrix}
          ,\quad
          \bv^{\prime} =
          \begin{pmatrix}
            0.1507  & -0.0505 & -0.1695 \\
            -0.0505 & 0.1601  & -0.1543 \\
            -0.1695 & -0.1543 & -0.3108
          \end{pmatrix},
        \end{equation*}
        and
        \begin{equation*}
          [\bH^{KS}_{TI}] =
          \begin{pmatrix}
            0.0176  & 0.0123  & -0.0299 & -0.0138 & -0.0969 & -0.0289 \\
            0.0123  & 0.0287  & -0.0409 & -0.0923 & -0.0195 & -0.0302 \\
            -0.0299 & -0.0409 & 0.0708  & 0.106   & 0.1165  & 0.0592  \\
            -0.0138 & -0.0923 & 0.106   & -0.0409 & 0.0592  & -0.0195 \\
            -0.0969 & -0.0195 & 0.1165  & 0.0592  & -0.0299 & -0.0138 \\
            -0.0289 & -0.0302 & 0.0592  & -0.0195 & -0.0138 & 0.0123
          \end{pmatrix}.
        \end{equation*}
\end{description}

\section{The generalized cross-product in components}
\label{sec:detailed-components}

The 10 independent components of the totally symmetric third order tensor $\ba \times \bb$, where both $\ba$ and $\bb$ are symmetric second order tensors, are:
\begin{equation}\label{eq:axbijk}
  \begin{aligned}
     & (\ba \times \bb)_{111} = a_{12} b_{13}-a_{13} b_{12},
    \\
     & (\ba \times \bb)_{112} = \frac{1}{3} (-a_{11} b_{13}+a_{12} b_{23}+a_{13} b_{11}-a_{13} b_{22}+a_{22} b_{13}-a_{23} b_{12}),
    \\
     & (\ba \times \bb)_{113} = \frac{1}{3} (a_{11} b_{12}-a_{12} b_{11}+a_{12} b_{33}-a_{13} b_{23}+a_{23} b_{13}-a_{33} b_{12}),
    \\
     & (\ba \times \bb)_{122} = \frac{1}{3} (-a_{11} b_{23}-a_{12} b_{13}+a_{13} b_{12}+a_{22} b_{23}+a_{23} b_{11}-a_{23} b_{22}),
    \\
     & (\ba \times \bb)_{123} = \frac{1}{6} (a_{11} b_{22}- a_{11}  b_{33}+a_{22} b_{33}-a_{22}b_{11}+a_{33} b_{11}-a_{33}b_{22}),
    \\
     & (\ba \times \bb)_{133} = \frac{1}{3} (a_{11} b_{23}-a_{12} b_{13}+a_{13} b_{12}-a_{23} b_{11}+a_{23} b_{33}-a_{33} b_{23}),
    \\
     & (\ba \times \bb)_{222} = a_{23} b_{12}-a_{12} b_{23},
    \\
     & (\ba \times \bb)_{223} = \frac{1}{3} (a_{12} b_{22}-a_{12} b_{33}-a_{13} b_{23}-a_{22} b_{12}+a_{23} b_{13}+a_{33} b_{12}),
    \\
     & (\ba \times \bb)_{233} = \frac{1}{3} (a_{12} b_{23}+a_{13} b_{22}- a_{13} b_{33}-a_{22} b_{13}-a_{23} b_{12}+a_{33} b_{13}),
    \\
     & (\ba \times \bb)_{333} = a_{13} b_{23}-a_{23} b_{13}.
  \end{aligned}
\end{equation}

For the components of its trace (which is a vector) one has
\begin{align*}
   & (\tr(\ba \times \bb))_{1}=\frac{1}{3} (a_{12} b_{13}-a_{13} b_{12}+a_{22} b_{23}-a_{23} b_{22}+a_{23} b_{33}-a_{33} b_{23}),
  \\
   & (\tr(\ba \times \bb))_{2}=\frac{1}{3} (-a_{11} b_{13}-a_{12} b_{23}+a_{13} b_{11}-a_{13} b_{33}+a_{23} b_{12}+a_{33} b_{13}),
  \\
   & (\tr(\ba \times \bb))_{3}= \frac{1}{3} (a_{11} b_{12}-a_{12} b_{11}+a_{12} b_{22}+a_{13} b_{23}-a_{22} b_{12}-a_{23} b_{13}).
\end{align*}

When $\bS$ is a totally symmetric fourth order tensor and $\ba$ is a symmetric second order tensor, the ten independent components of the totally symmetric third order tensor $\tr(\bS \times \ba)$ are
\begin{equation}\label{eq:TrSxaijk}
  \small
  \begin{aligned}
    (\tr(\bS \times \ba))_{111} = & \frac{3}{10} \Big(-a_{12} (2 S_{1113}+S_{1223}+S_{1333})+a_{13} (2 S_{1112}+S_{1222}+S_{1233})-a_{22} S_{1123}
    \\
                                  & +a_{23} (S_{1122}-S_{1133})+a_{33} S_{1123}\Big),
    \\
    (\tr(\bS \times \ba))_{112} = & \frac{1}{10} \Big(a_{11} (2 S_{1113}+S_{1223}+S_{1333})- a_{12} (2S_{1123}+S_{2223}+ S_{2333})
    \\
                                  & +a_{13} (-2 S_{1111}+2 S_{1122}+S_{2222}+S_{2233})-a_{22} (S_{1113}+3 S_{1223}+ S_{1333})
    \\
                                  & +a_{23} (3 S_{1222}- S_{1233})+ a_{33} (2S_{1223}-S_{1113})\Big),
    \\
    (\tr(\bS \times \ba))_{113} = & \frac{1}{10} \Big(-a_{11} (2 S_{1112}+S_{1222}+S_{1233})+a_{12} (2 S_{1111}-2 S_{1133}-S_{2233}-S_{3333})
    \\
                                  & +a_{13}(2  S_{1123}+S_{2223}+S_{2333})+a_{22} (S_{1112}-2 S_{1233})+a_{23} (S_{1223}-3  S_{1333})
    \\
                                  & +a_{33} (S_{1112}+ S_{1222}+3 S_{1233})\Big),
    \\
    (\tr(\bS \times \ba))_{122} = & \frac{1}{10} \Big(a_{11} (3 S_{1123}+S_{2223}+ S_{2333})+a_{12} (S_{1113}+2 S_{1223}+ S_{1333})+a_{13} (S_{1233}-3 S_{1112})
    \\
                                  & -a_{22}( S_{1123}+2 S_{2223}+ S_{2333})-a_{23} (S_{1111}+2 S_{1122}+S_{1133}-2 S_{2222})+a_{33} (S_{2223}-2 S_{1123})\Big),
    \\
    (\tr(\bS \times \ba))_{123} = & \frac{1}{20} \Big( a_{11} (3S_{1133}-3S_{1122}-S_{2222}+S_{3333})+2a_{12} ( S_{1112}- S_{1222})+2 a_{13} (S_{1333}-S_{1113})
    \\
                                  & +a_{22} (S_{1111}+3 S_{1122}-3 S_{2233}-S_{3333})+2 a_{23}( S_{2223}- S_{2333})
    \\
                                  & +a_{33} (-S_{1111}-3 S_{1133}+S_{2222}+3 S_{2233})\Big),
    \\
    (\tr(\bS \times \ba))_{133} = & \frac{1}{10} \Big(-a_{11}(3  S_{1123}+ S_{2223}+ S_{2333})+a_{12} (3 S_{1113}- S_{1223})-a_{13} (S_{1112}+S_{1222}+2 S_{1233})
    \\
                                  & +a_{22} (2 S_{1123}-S_{2333})+a_{23} (S_{1111}+S_{1122}+2 S_{1133}-2 S_{3333})+a_{33} (S_{1123}+S_{2223}+2 S_{2333})\Big),
    \\
    (\tr(\bS \times \ba))_{222} = & \frac{3}{10} \Big(a_{11} S_{1223}+a_{12} (S_{1123}+2 S_{2223}+ S_{2333})+a_{13} (S_{2233}-S_{1122})
    \\
                                  & -a_{23} (S_{1112}+2 S_{1222}+S_{1233})-a_{33} S_{1223}\Big),
    \\
    (\tr(\bS \times \ba))_{223} = & \frac{1}{10} \Big(a_{11}(2  S_{1233}- S_{1222})+a_{12} (S_{1133}-2  S_{2222}+2  S_{2233}+S_{3333})+a_{13}(3  S_{2333}- S_{1123})
    \\
                                  & +a_{22} (S_{1112}+2 S_{1222}+S_{1233}) -a_{23} (S_{1113}+2 S_{1223}+S_{1333})-a_{33} (S_{1112}+S_{1222}+3 S_{1233})\Big),
    \\
    (\tr(\bS \times \ba))_{233} = & \frac{1}{10} \Big(a_{11} ( S_{1333}-2 S_{1223})+a_{12} (S_{1123}-3  S_{2223})+a_{13} (2 S_{3333}-S_{1122}- S_{2222}-2  S_{2233})
    \\
                                  & +a_{22} (S_{1113}+3 S_{1223}+S_{1333})+a_{23} (S_{1112}+S_{1222}+2 S_{1233})-a_{33} (S_{1113}+ S_{1223}+2 S_{1333})\Big),
    \\
    (\tr(\bS \times \ba))_{333} = & \frac{3}{10} \Big(-a_{11} S_{1233}+a_{12} (S_{1133}-S_{2233})-a_{13} (S_{1123}+S_{2223}+2 S_{2333})+a_{22} S_{1233}
    \\
                                  & +a_{23} (S_{1113}+S_{1223}+2 S_{1333})\Big).
  \end{aligned}
\end{equation}


\end{document}